# The Landscape of Undergraduate Astronomy and Astrophysics Degree Requirements

*Delivered to AAS Board of Trustees, December 22, 2025*

By the AAS Education Committee Subcommittee on UndeRgraduate and Graduate Education (SURGE)
Kate Follette (Amherst College, chair)
Carl Ferkinhoff (Winona State University)
Michael Foley (Harvard University)
Meridith MacGregor (Johns Hopkins University)
Melissa Morris (Lycoming College)
Karen Masters (Haverford College)
Tom Rice (AAS Education Officer)
Colin Wallace (University of North Carolina)

*This report and its recommendations have been endorsed by: the AAS Education Committee, Employment Committee, Working Group on Graduate Admissions, and Beyond Academic Careers Advisory Committee*

# Introduction and Overview

25 years ago, the typical education path for a professional astronomer lay through an undergraduate degree in physics, followed by specialized PhD study in astronomy or astrophysics. At that time, the number of undergraduate degrees awarded in Astronomy/Astrophysics had remained steady at ~200 per year since the 1970s (see Fig. 1). In the past two and a half decades, however, the number of undergraduate degrees in Astronomy and Astrophysics has skyrocketed, as has the number of programs offering such degrees. In 2024, 982 students completed undergraduate degrees in Astronomy and Astrophysics in the United States (AIP 2025), a fivefold increase since the year 2000.

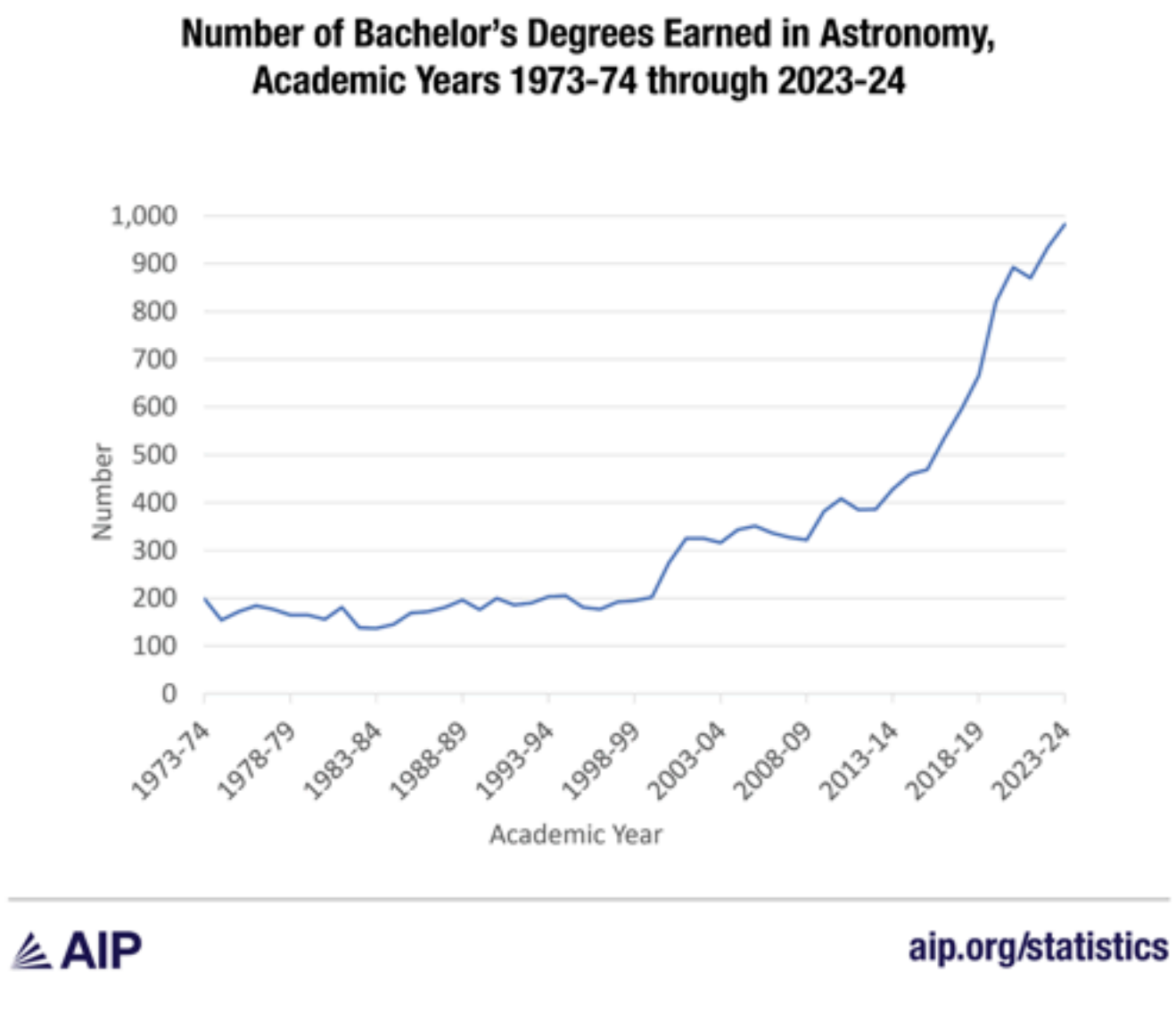

**Figure 1.** American Institute of Physics enrollment statistics for undergraduate majors in astronomy and astrophysics from 1973 to 2024, showing a steep rise since ~2000, and an even steeper rise in the past decade.

The relative novelty of specialized bachelors' programs in Astronomy and Astrophysics means that little systemic coordination or study of them has yet been done. Is there any consistency in what is expected from graduates of undergraduate Astronomy/Astrophysics degree programs? Do students, faculty, and prospective employers have a shared understanding of the skills and knowledge that such programs instill? Motivated by these questions and those in Figure 2 below, we on the American Astronomical Society's Education Committee Subcommittee on UndeRgraduate and Graduate Education (hereafter SURGE) conducted a study of undergraduate degree-granting programs in the United States. Our survey was administered from February through June of 2025 and was completed by 88 individuals representing 78 unique departments. This report reflects findings and recommendations resulting from our analyses of these data.

**Summary of the AAS Education SURGE Survey of Undergraduate Degree-Granting Programs**

**Goal:** Compile and analyze information that will allow our subcommittee, the AAS Education committee, and AAS leadership to better understand the current landscape of *undergraduate* Astronomy and Astrophysics programs. This effort is designed to provide us with the information that we need to support such programs, establish communities of support, and prioritize change efforts.

Our efforts are particularly motivated by the following questions:

1. What constitutes an Astronomy/Astrophysics major? What elements are shared across programs? How much variety is there?
2. What are departments preparing their students ***for*** with Astronomy or Astrophysics majors? Graduate school? Jobs in industry? Education/outreach?
3. What can/should an undergraduate Astronomy/Astrophysics major communicate to graduate admissions committees and prospective employers? (i.e. what's the point?)
4. What is the basic set of skills and knowledge that all Astronomy/Astrophysics majors share?
5. How do undergraduate programs' goals and needs vary?

**Fig. 2.** Summary of the goals and foundational questions underlying our survey effort, drafted as we began this project and serving as a guide throughout.

Our report begins in section 1 with a brief description of how the survey was developed and administered. The full text of the survey, as well as detailed analyses of each of its questions, is provided in Appendix A of this report. Survey data products that we believe will prove valuable to various AAS committees and other interested groups are linked in Appendix A. In Section 2, we present our top-level findings and recommendations, and conclude briefly in Section 3.

# 1. Membership and Workflow

## *1.1 Committee Membership*

The survey of undergraduate degree-granting programs was conducted by the 2024-2025 American Astronomical Society Education Committee's Subcommittee on UndeRgraduate and Graduate Education (SURGE) in its inaugural year as a standalone entity. Several members (Follette, Ferkinhoff, Foley, and MacGregor) were assigned to the subcommittee in their capacity as members of the broader AAS Education Committee, and several others (Masters, Morris, Wallace) volunteered for the subcommittee as affiliate members. Tom Rice, the AAS Education Officer, was actively involved in all stages of the subcommittee's work. We note that our members span a range of institutional environments and career stages, and all have contributed to and endorsed the findings and recommendations presented below, which speaks to their broad utility and applicability across institutional realities.

## *1.2 Workflow*

The committee's work was conducted in three phases over the course of one year, from September 2024 to August 2025. The committee met biweekly from September through December, and again from March through July, 2025.

***Phase 1: Brainstorming*** . The committee spent much of fall 2024 brainstorming ideas for the survey, with a particular focus on defining its scope, focus, and goals, as summarized in Figure 2. The impetus to design the survey was provided in part by the AAS Task Force on Early Career Engagement's

Recommendation 1b, namely: *“The AAS Education Committee and Beyond Academic Careers Advisory Committee should work together to develop a set of recommendations for undergraduate and graduate curricula that adequately prepare students for careers in both academia and industry.”* In order to fulfill this recommendation, our committee determined that we needed first to understand the landscape of undergraduate Astronomy curricula across the country. Our thinking was also influenced by debates and problems that have arisen at recent AAS special sessions organized by the Education Committee co-chairs, especially the Winter 2024 meeting’s “How We Move Forward in Supporting Students in the Field” organized by former AAS Education Committee co-chairs Sanlyn Buxner and Karen Masters.

***Phase 2: Survey Drafting*** took place from November-December of 2024. The final product was a 39 question survey targeting departments with degrees requiring astronomy or astrophysics coursework, but including another (much shorter) path for those without Astronomy/Astrophysics degrees (see Appendix A for the full text of the questions). The final survey was distributed through the AAS Education blog, AAS Education newsletter, and the AAS departmental chairs list, as well as being advertised at the AAS 245 meeting in various ways. Following initial collection of survey responses, SURGE members conducted targeted outreach to an additional 48 programs that were missing from the dataset, of which 27 ultimately completed the survey.

***Phase 3: Survey Analysis and Drafting of Findings and Recommendations*** took place from March to July, 2025. Individual Task Force members conducted quantitative analyses of responses to the multiple choice questions on the survey and qualitative thematic analyses of the free response questions. Graphical, tabular, and written summaries of these analyses are outlined in detail in Appendix A of this report, which also links the raw data products. While analyses were conducted initially by individual or small groups of committee members, all results were presented to and iterated upon by the full committee and served as the basis for the findings and recommendations below, which were drafted and reviewed collaboratively.

# 2. Findings and Recommendations

***Those wishing to understand more of the motivation behind the findings and recommendations listed below should visit the relevant section of Appendix A for more information and detailed analyses of survey responses.*** Clickable links to each Appendix Section, which correspond to the types of questions asked on our survey, appear below

We note that these findings and recommendations reflect the thoughts and opinions of the members of our committee based on our reading of the survey results and our own experiences. We view this report as only the beginning of an important dialog, and we invite commentary via this google form.

## 2.1 Key Findings

From among our extensive analyses of survey data, the committee selected the following nine key findings as especially significant, surprising, or important. Please note that many additional results, as well as additional detail about the analyses leading to these findings, can be found in Appendix A.

***Finding 1:*** Opportunities to major or concentrate in Astronomy/Astrophysics span many types and sizes of department and institution, and are offered as BA and BS degrees in nearly equal proportion (49.6% and 50.4%, respectively) .

***Finding 2:*** Only 34% of the departments surveyed exist as separate Departments of Astronomy and/or Astrophysics. Most Astronomy/Astrophysics degrees are offered in departments with "physics" in the name (43% in Departments of Physics and Astronomy and 15% in Departments of Physics).

***Finding 3:*** 85% of departments that completed the survey offered an undergraduate degree or concentration in Astronomy and/or Astrophysics, representing a total of 66 departments and an estimated total of 718-1007 undergraduate majors annually. This is in keeping with current AIP statistics (982 bachelor's degrees awarded in the 2023-24 academic year), indicating that we have captured a large fraction of all undergraduate degree-granting programs in the United States.

***Finding 4:*** There are a roughly equivalent number of degrees titled "Astronomy" (36%) and "Astrophysics" (30%) among the survey sample. The number of courses required for "Astrophysics" majors is, on average, *slightly* higher (21±6) than the number of courses required for "Astronomy" majors (18±5). Similarly, the number of courses required for a BS (22 ±5) is *slightly* higher than for a BA (15 ± 3). However, neither difference was significant.

***Finding 5:*** Although "Astronomy" and "Astrophysics" majors across the country do not differ significantly in the total number of courses that they require, the terms have different popular interpretations, with "astrophysics" being perceived as more rigorous than "astronomy".

***Finding 6:*** Differential calculus, integral calculus, introductory mechanics, and introductory electricity and magnetism are the ***only*** courses required by ***all*** Astronomy/Astrophysics degree programs.

***Finding 7:*** Roughly ⅔ of surveyed departments indicated that their department has formal learning goals for their Astronomy/Astrophysics major. A majority (52%) indicated that their department analyzes progress toward those learning goals at least annually.

***Finding 8:*** Survey responses show broad agreement on the core competencies that undergraduate Astronomy and Astrophysics majors should develop, but also reveal gaps between perceptions of ideal program outcomes ("informal" learning goals) and formal learning objectives. Fourteen shared themes emerged from our analysis of informally- and formally-expressed learning goals. Listed in order of their prevalence among formal learning goals, these are: (1) Physics Knowledge and Skills, (2) Problem Solving or Critical Thinking Skills, (3) Communication Skills, (4) Observational, Experimental, Instrumentation or Laboratory Skills, (5) Astronomy and Astrophysics Knowledge and Skills, (6) Data, Computational, and Programming Skills, (7) Mathematics, Statistics, or Quantitative Knowledge and Skills, (8) Scientific Reasoning, Method, or Process Skills, (9) Research Experience and Skills, (10) Ethics and Inclusion Training or Skills, (11) Training for Industry, Careers, or the Profession Broadly, (12) Preparation for Graduate Work, (13) Collaboration Skills, and (14) Reading Academic Literature or Finding Scientific Information. The same key skills or competencies emerge from analyses of informally-expressed learning goals, but differ markedly in the rates at which such skills are mentioned, suggesting opportunities to better integrate formal program-level learning goals with field-based expectations and norms.

***Finding 9***: For most programs, there is a wide range in the rates of student engagement in research and plans to attend graduate school. In other words, Astronomy and Astrophysics programs contain a highly heterogeneous mix of student experiences and goals. It is rare for a department to consist of students with homogeneous interests and paths.

## *2.2 Our Recommendations*

The recommendations below represent the consensus opinion of SURGE, and thus reflect the wide array of experiences and institutional environments of our members. They have been endorsed by the full AAS Education and Employment Committees, as well as the Working Group on Graduate Admissions and the Beyond Academic Careers Advisory Committee, which represent an even broader array of contexts. ***However, we recognize that the conclusions we have drawn from our data cannot account for the institutional and political realities of every program. They are made with the intention of prompting reflection within programs, rather than mandating or providing a precise recipe for change.***

**Recommendation 1:** Astronomy/Astrophysics majors show healthy enrollments across a wide range of institution, department, and degree types, revealing a broad appetite for this specialization at the undergraduate level. As such, we recommend that departments that do not currently offer an astronomy-specific credential adopt a major or concentration in Astronomy/Astrophysics.

**Recommendation 2:** As our data do not support a distinction between “Astronomy” and “Astrophysics” degrees, as this variance in terminology promotes a false sense of difference, and as our understanding is that a major with “physics” in the name allows for greater career opportunities for graduating students, we recommend that departments adopt the term “astrophysics” to describe their academic majors *wherever it is feasible to do so*.

**Recommendation 3:** Our data show that currently the **only** shared requirements of all Astronomy and Astrophysics programs are (a) two semesters of calculus and (b) two introductory physics courses. This is in stark contrast to relatively homogeneous undergraduate programs in most other STEM disciplines, including physics. So that a degree in Astronomy or Astrophysics communicates a basic, shared set of knowledge and skills that transcends institution, we recommend that the AAS Board of Trustees endorse a set of *recommended* course requirements for Astronomy/Astrophysics undergraduate degree-granting programs. Based on common course requirements across the 78 programs we surveyed, our own experiences as educators, and consultation with the AAS Employment Committee, Working Group on Graduate Admissions, and Beyond Academic Careers Advisory Committee, ***we propose in Appendix B a possible undergraduate course requirement list*** with the aim of providing a starting point for this important conversation.

**Recommendation 4:** We recommend that all academic departments adopt formal program-level learning goals that are well-aligned with expected outcomes for students graduating from the program. These should encompass the content knowledge and skills that they believe ***all*** majors should leave the program with. These learning goals should map to specific (ideally multiple) courses within the major requirements and should be phrased in such a way that they can be easily assessed, both by the department/institution internally and externally by departmental reviewers or accreditors. ***Appendix C provides examples of well-phrased program-level learning goals culled from our data and representing each of the main shared themes from our analyses.***

**Recommendation 5:** Given the high rate at which astronomers who responded to our survey specified data analysis and scientific computing skills among the most important things majors should leave Astronomy/Astrophysics programs with, we especially strongly recommend that all programs that do not currently have formal learning goals in this area adopt them.

**Recommendation 6:** Modern astronomers rely heavily on programming and computation, and these are highly transferable skills that Astronomy/Astrophysics majors can leverage to seek careers beyond academia, which a majority of them do post-graduation. Therefore, we recommend that departments work to embed computational competencies as early as possible into the coursework for Astronomy and Astrophysics majors. We recommend that the AAS and its affiliate committees place a high priority on the development of pedagogical resources aimed at teaching scientific computing in Astronomy.

**Recommendation 7:** That the AAS sponsor a virtual or in-person meeting of the existing "Astronomy Department Chairs and Program Directors" group with the goal of revising as necessary and endorsing the recommendations provided in this section, and in Appendices B and C of this report. Given the high proportion of Astronomy/Astrophysics degrees offered in Departments of Physics, we recommend targeted outreach inviting such departments to send a representative.

**Recommendation 8:** That the AAS create and maintain its own email list of chairs of departments offering undergraduate and/or graduate programs in Astronomy/Astrophysics. This list should be used sparingly to support transmission of key information from AAS committees and leadership that is especially relevant at the department level.

**Recommendation 9:** PhD programs in Astronomy and Astrophysics now admit roughly 200 students per year, approximately double the number they admitted in the year 2000, but still only ~20% of the number of undergraduate majors in the discipline. It is clear from these data, and from our survey responses, that undergraduate Astronomy programs ***must*** prepare students for a wide array of careers. Care should be taken to ensure that any course recommendations and learning goals endorsed by the AAS for Astronomy/Astrophysics majors are relevant across the broad spectrum of careers that Astronomy/Astrophysics majors pursue, and are not narrowly focused on graduate preparation.

**Recommendation 10:** The AAS Working Group on Graduate Admissions should work to develop a more complete list of required/recommended knowledge and skills/competencies for graduate school and advertise it broadly to both undergraduate programs and early career researchers. *We do not endorse a move toward all undergraduate astronomy curricula fulfilling all grad school requirements* (see recommendation 9), but believe that undergraduate programs should consider these needs in planning and advertising their curricula and elective course offerings, as well as advising undergraduates who intend to pursue graduate studies.

# 3. Summary and Conclusion

The survey results highlighted in Section 2 and described in detail in Appendix A have answered some of the questions that we sought to answer through this effort, elucidated the complexities of others, and raised several altogether new ones. We, the authors of this report, believe that a much broader swath of the community needs to engage in discussion around undergraduate Astronomy and Astrophysics curricula. In addition to our findings and recommendations above, our detailed analyses in Appendix A, and the public dataset linked there, we would like to leave the community with the following five questions. While acknowledging that much work remains to be done, we hope that our report has substantively informed several of them.

1. What is the difference between "Astronomy" and "Astrophysics"? What does each term communicate to students and to potential employers? Which term should be adopted and under what circumstances?
2. What coursework *should* constitute an Astronomy/Astrophysics major?
3. What *should* departments prepare their students for with Astronomy or Astrophysics majors? Graduate school? Jobs in industry? Education/outreach?
4. What is the basic set of skills and knowledge that all Astronomy/Astrophysics majors *should* have?
5. How do undergraduate programs' goals and needs vary?

# Appendix A

This appendix contains survey data analyses conducted by SURGE members. The analyses reported here formed the basis for our top-level findings and recommendations, as described in Section 2 of this report. Drafting of each section of this Appendix was led by a small group of committee members, and audited by at least one additional member. All analyses were presented to the SURGE committee for comment, and revised. Survey data products are available here. The data are separated into the following tabs.

1. "Clean Data" contains one response for each unique institution represented in the dataset
2. "All Data" contains the raw survey responses, with duplicates (multiple responses from the same institution) highlighted
3. "Duplicate analysis" contains the responses for each duplicated institution and a simple analysis of whether the responses agree with one another
4. The "retention rates" and "other" tabs contain scrambled responses to the two survey questions for which participants were told their responses would not be identifiable
5. "Learning Goal Analysis" contains the results of a qualitative thematic coding analysis of free response questions concerning departmental learning goals

All survey data are provided in their unaltered format except in cases where we felt that a response to one of the two anonymous questions (item 4 above) compromised the anonymity of the respondent (N=11), in which case we redacted certain identifiable phrases.

Each section below begins with a table listing the full text of the questions and available responses in that section, as well as text describing who was asked the question(s). This is followed by a "Summary and Key Takeaways" subsection with a high-level summary of our analyses. The final subsection lists quantitative and qualitative (thematic) analyses[1] for each question and a summary of our interpretation.

The survey preamble read:

> We are compiling information about the landscape of Astronomy and Astrophysics ***undergraduate*** programs in the United States. Our goal is to understand the breadth and depth of undergraduate curricula and degree requirements nationally. We aim to use this information to compile a report about the landscape of undergraduate Astronomy and Astrophysics majors. You may indicate at the end of the survey whether you would like to receive a copy of that report. No institutions will be singled out by name in that report.
>
> **While we recognize that many astronomers receive degrees in physics, planetary science, etc., for the purposes of this study, we are specifically interested in degrees that specifically require Astronomy and/or Astrophysics coursework.**
>
> We anticipate that the survey should take you approximately 5-10 minutes to complete. Thank you very much for your time in supporting this effort.

[1] We are not experts in qualitative data analyses. Other interested parties may identify additional or different themes in the free response data, which is why we provide it together with this report.

# 1. Information About Participating Departments

This section consisted of 7 of questions, summarized below

| No | Question Category | Who Asked | Question Text | Answer Text<br>(Display Logic) |
|---|---|---|---|---|
| 1 | Dept. Info | All | What is the name of the **institution** whose undergraduate program you will be answering these questions about?<br>*ideally an institution where you currently work or are a student* | Free response |
| 2 | Dept. Info | All | What is the name of the **department** whose undergraduate program you will be answering these questions about (*hereafter "your department"*)? | Department of Physics<br>Department of Astronomy<br>Department of Physics and Astronomy<br>Other |
| 3 | Dept. Info | All | What is **your role** within that department? (Choose all that apply) | Department Chair<br>Undergraduate program head<br>Faculty member<br>Current student<br>Former student<br>Other |
| 4 | Dept. Info | All | How would you **classify your institution** from among the following choices? Please choose all that apply. | Community college<br>Liberal arts college<br>Private research university (grants doctoral degrees)<br>Public research university (grants doctoral degrees)<br>Regional comprehensive college (grants masters degrees)<br>Hispanic-serving institution<br>Historically Black College or University (HBCU)<br>Religious college<br>Tribal college<br>Women's college<br>Other |
| 5 | Dept. Info | All | What **style of academic terms** do you have at your institution? | Semesters<br>Trimesters<br>Quarters<br>Other |
| 6 | Dept. Info | All | Which of the following **graduate degrees** does your department/institution offer ***in Astronomy and/or Astrophysics***? Please choose all that apply. | A Master's degree (MS)<br>A Doctoral degree (PhD)<br>We do not offer higher degrees |
| 7 | Dept. Info | All | How many **types** of **UNDERGRADUATE degrees** in Astronomy or Astrophysics does your department offer? | One<br>More than one (e.g. both "Astronomy" and "Astrophysics", both a BA and a BS) → Q12<br>We do not offer degrees or concentrations in Astronomy or Astrophysics → Q17 |

## *1.1 Summary and Key Takeaways*

We received responses from a total of 78[2] institutions. As stated in Finding #3 above, the vast majority (85%) of the institutions responding to our survey offer an undergraduate degree or concentration in Astronomy and/or Astrophysics, with 49% offering more than one type of degree in the field (e.g. a BA and a BS, or a degree in Astronomy and another in Astrophysics, see Section 2 for further detail). Institutions offering undergraduate degrees in the field span a wide range, including 26% that serve undergraduates only and do not offer higher level degrees, and an additional 20% that do not offer higher degrees in Astronomy or Astrophysics.

The survey was mostly completed by faculty, department chairs, or undergraduate program heads (83% of respondents), but also included a small number of current or former students (11.4 and 5.7% of the 88 survey respondents, respectively). Nearly all of the responding institutions (92%) are on a semester system. Figures 3 and 4 report statistics on department name and type of institution, respectively. Appendix D lists all institutions that responded to the survey.

## *1.2 Data Summary and Analysis*

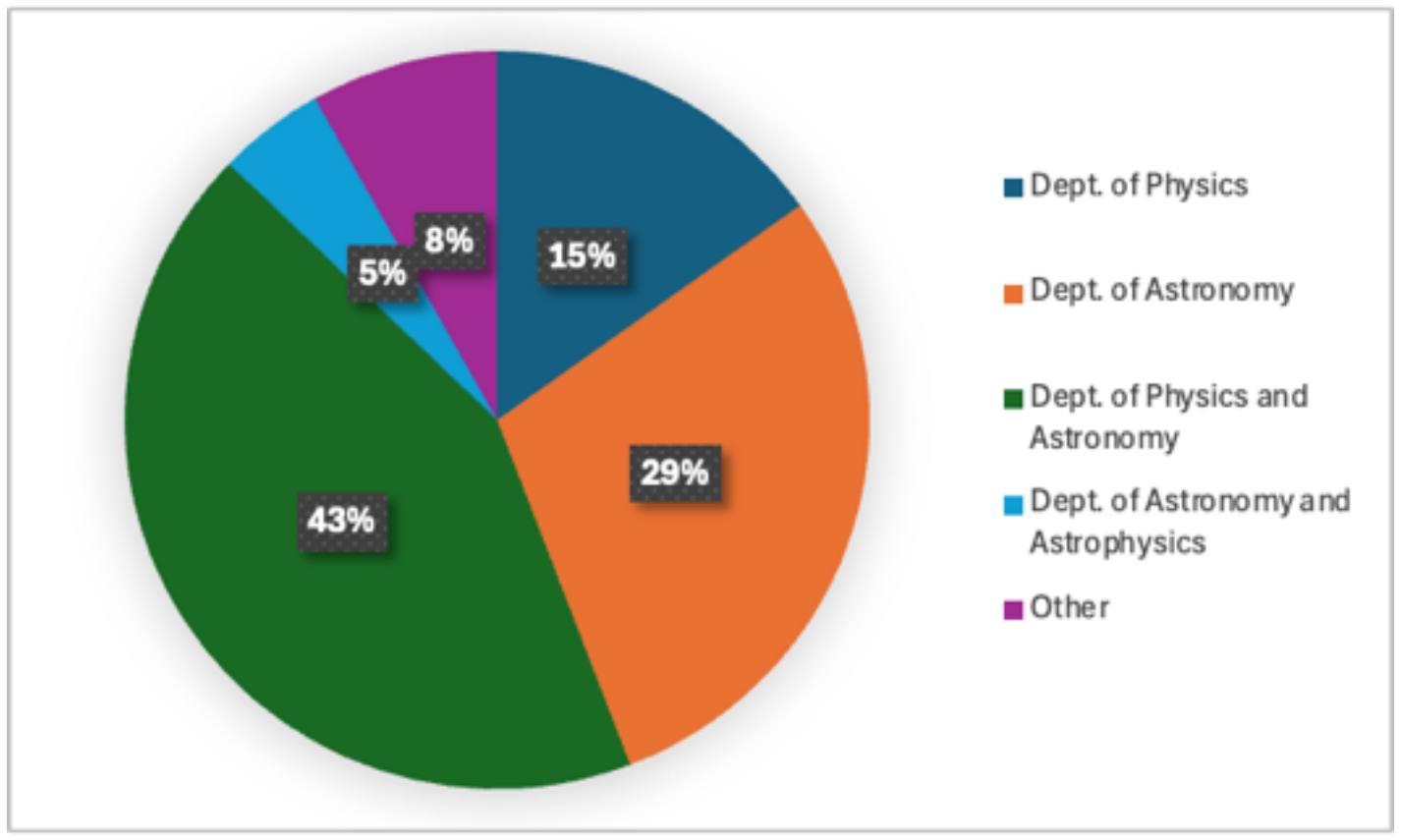

**Figure 3:** Names of departments among institutions that completed the survey.

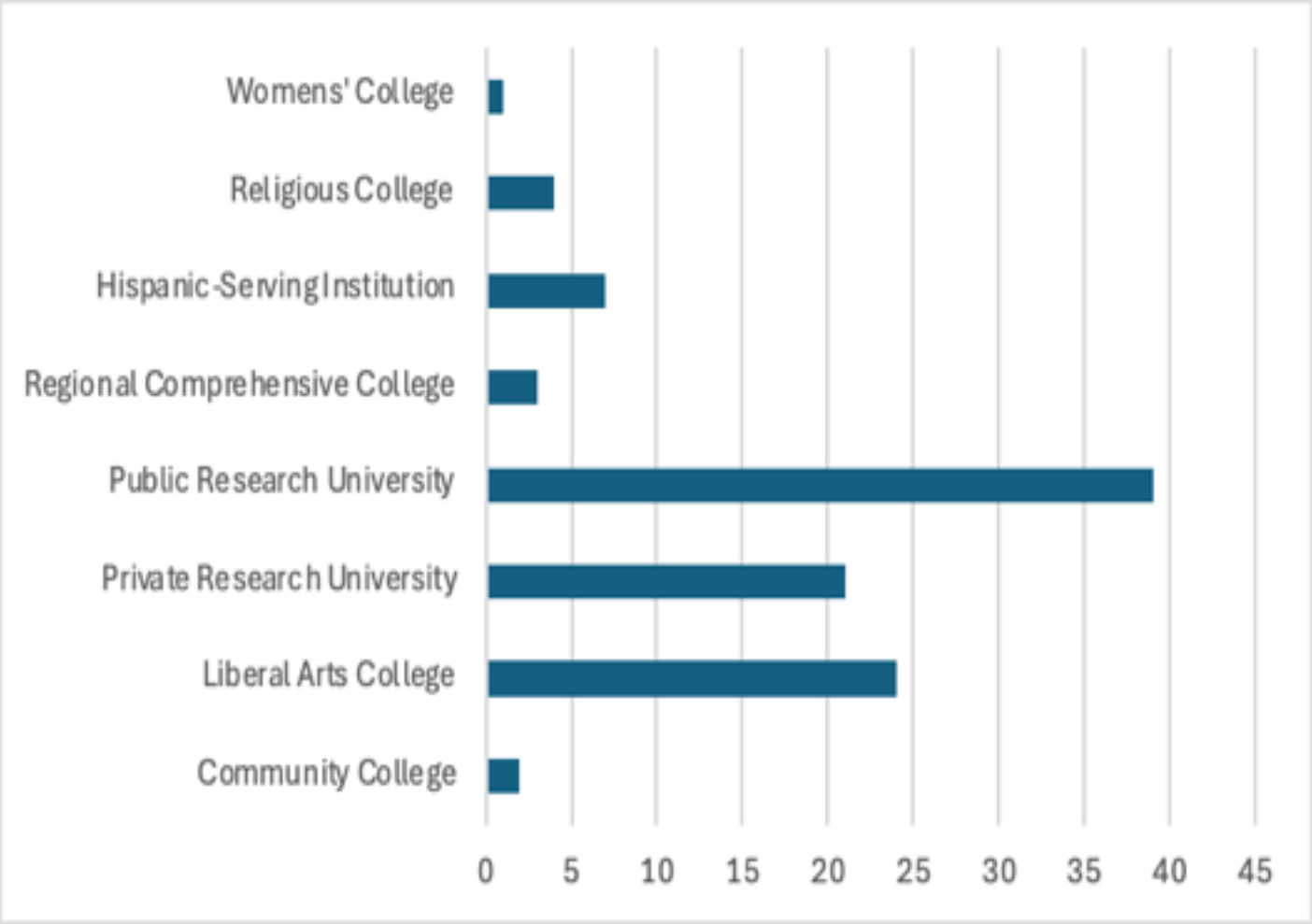

**Figure 4:** Types of departments that responded to the survey. Participants were instructed to choose all that applied.

---

[2] One additional institution was removed from the raw dataset. The survey respondent, a current student, indicated that their institution did not have an Astronomy major, but a check of the institution's website indicated that this was not correct, so the response was removed.

# 2. Astronomy and Astrophysics Degrees

Respondents to the survey encountered three different versions of questions about their Astronomy and/or Astrophysics degree offerings, depending on whether they indicated that their department offered one, more than one, or no undergraduate degrees in these subjects.

Departments offering one degree were asked 4 questions (Q8-11), departments offering multiple degrees were asked 5 questions (Q12-16), and departments not offering any degrees in Astronomy and/or Astrophysics were asked 2 questions (Q17-18).

| No | Question Category | Who Asked | Question Text | Answer Text (Display Logic) |
|---|---|---|---|---|
| 8 | Degree Info | Single degree depts | What is the **name** of the undergraduate Astronomy or Astrophysics degree that your department offers? | "Astronomy"<br>"Astrophysics"<br>"Astronomy and Astrophysics"<br>"Physics and Astronomy"<br>Physics with a concentration in "Astronomy" or "Astrophysics"<br>Other |
| 9 | Degree Info | Single degree depts | What is the **type** of the undergraduate Astronomy or Astrophysics degree that your department offers? | A Bachelor of Arts degree (BA)<br>A Bachelor of Science degree (BS)<br>Other |
| 10 | Degree Info | Single degree depts | On average, approximately **how many** of these undergraduate degrees does your department award annually?<br>*you may also give a range (e.g. 5-10 majors per year)* | Free response |
| 11 | Degree Info | Single degree depts | How many **total** courses are required for an undergraduate to complete this major?<br>*Note: courses that count towards the major only, NOT general education or distribution requirements imposed by the university.*<br>*If able, please specify both the number of courses that formally comprise the major AND the number of additional courses required as prerequisites to enroll in the major's required courses (e.g. calculus, if not formally "counted" toward the major*<br>*For example:*<br>*"We require 11 courses in physics and astronomy to complete the major. 3 additional math and 1 computer science course are required*<br>*as prerequisites to enroll in those courses"* | Free response → Q19 |
| 12 | Degree Info | Multiple degree depts | What are the **name(s)** of the undergraduate Astronomy or Astrophysics degree(s) that your department offers? Please choose all that apply. | Same as Q8 |
| 13 | Degree Info | Multiple degree depts | What are the **types** of undergraduate Astronomy or Astrophysics degree that your department offers? Please choose all that apply. | Same as Q9 |
| 14 | Degree Info | Multiple degree depts | On average, approximately **how many of each type** of undergraduate degree does your department award annually?<br>*you may also give a range (e.g. 5-10 majors per year)* | Free response |
| 15 | Degree Info | Multiple degree depts | If your department offers more than one undergraduate degree in astronomy or astrophysics, we ask you to complete the remainder of this survey for ***only your most popular*** | Free response |

| No | Question Category | Who Asked | Question Text | Answer Text (Display Logic) |
|---|---|---|---|---|
| | | | ***astronomy/astrophysics major.***<br><br>Please indicate which major that is in the space below and describe the basic differences between it and your other degrees<br><br>*For example: I will answer for our BS degree in Astrophysics. Our BA in Astronomy requires 9 courses total, 3 fewer physics courses than*<br>*the BS.* | |
| 16 | Degree Info | Multiple degree depts | Same as Q11 ("How many **total** courses…") | Free response → Q19 |
| 17 | Degree Info | No degree depts | Has your department considered offering an undergraduate degree in Astronomy or Astrophysics? | Yes<br>No<br>I don't know<br>Other |
| 18 | Degree Info | No degree depts | How does your department currently meet the educational needs of students interested in Astronomy and Astrophysics? | Free response → Q39 |

## *2.1 Summary and Key Takeaways*

In our analysis, we consolidated responses from programs with multiple survey respondents (n=9), checking that both respondents' answers agreed with one another within reason. Where possible, we prioritized the responses of faculty members over students, based on the assumption that faculty would be most familiar with the curriculum and most likely to answer accurately. In the end, there were 88 respondents from 78 different institutions. Of these, 66 institutions reported having an Astronomy or Astrophysics-related major or concentration.

Among all institutions offering an Astronomy or Astrophysics-related major, there was a 49.6%-50.4% split between BAs and BSes offered. 37 of the 66 Departments (56%) that offered an Astronomy/Astrophysics specialization specified that they offer more than one degree type. The majority of these programs offered both a BS and a BA degree, with some explicitly commenting that the BS is intended for those planning on going to grad school. Of the programs that offered both a BS and a BA degree option, 95% reported that the BS was the more popular option.

Only 12 of the 78 surveyed programs (15%) indicated that they don't offer any Astronomy/Astrophysics major or concentration. Of these, roughly half said they have considered offering one. 7 of these departments indicated that they offer a minor in Astronomy or Astrophysics, while others report offering research experiences and/or significant electives (counting towards a physics major) in astronomical subjects. Only one survey respondent said there was zero opportunity for astronomically interested students at their institution. We note that this low rate of institutions without a major or concentration could reflect a bias in who received and/or responded to the survey, with institutions without Astronomy/Astrophysics programming being much less likely to respond.

## *2.2 Data Summary and Analysis*

### *2.2.1 Degree Names and Types*

A summary of the results to questions regarding degree names and types can be found in Figure 5.

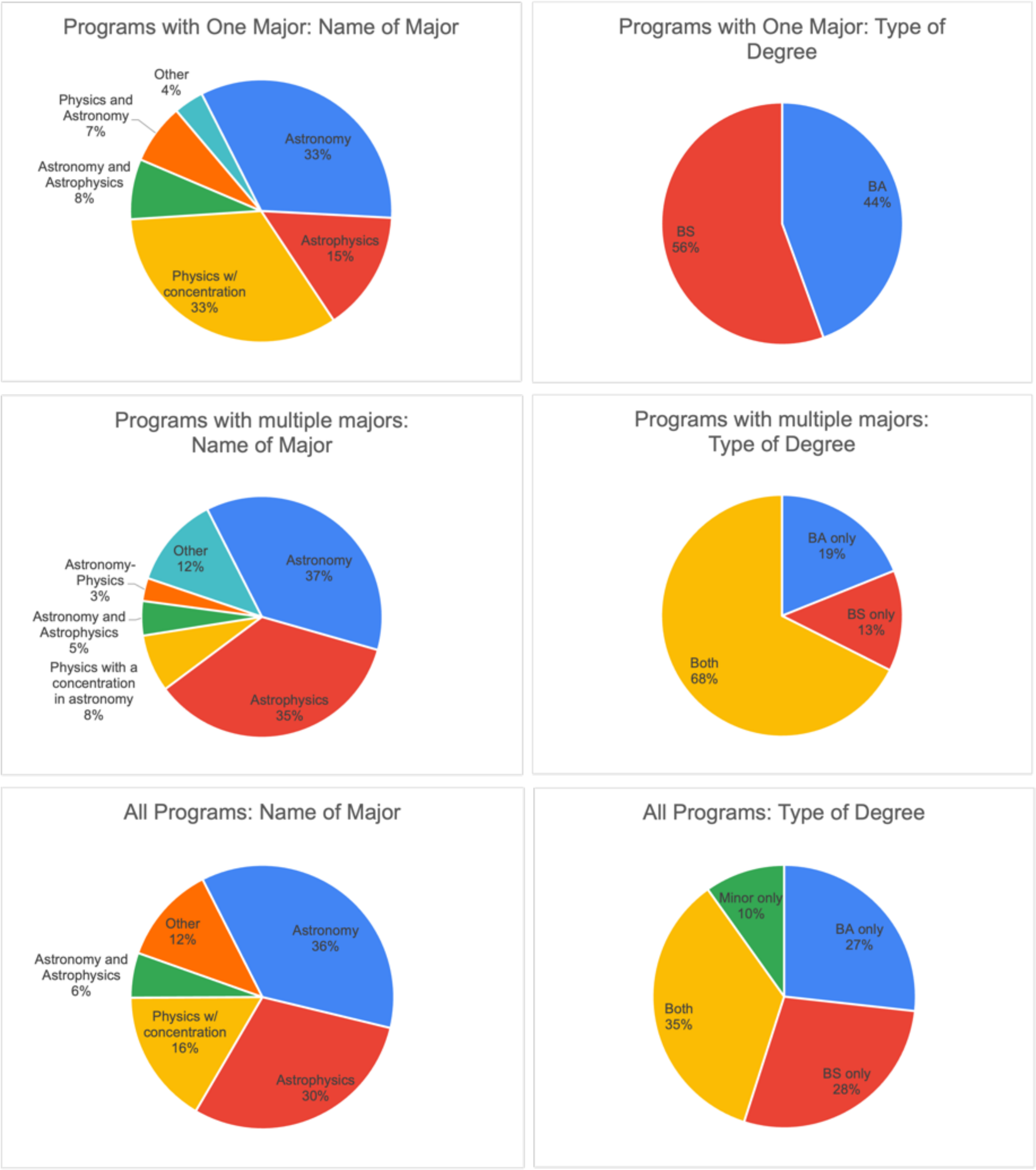

**Figure 5:** The variation in names of majors (left column) and types of degrees (right column) offered, split by programs that offer one major (top row), programs that offer multiple majors (middle row), and all programs combined (bottom row).

Upon examining responses to these particular questions, we find some differing trends between institutions that offer only one Astronomy or Astrophysics-related degree and those that offer multiple.

Of the institutions offering one degree, there is a slight preference towards a BS (56%) degree in the survey sample. Names are split, with roughly a third of the majors named “Astronomy”, a third

“Physics with a concentration in Astronomy or Astrophysics”, and the remaining split between “Astrophysics” and a variety of other names.

For institutions that offer multiple degrees, the majority (68%) offer both a BA and a BS degree. Of the remainder, slightly more offer two BA degrees than two BS degrees. Most degrees in dual-degree departments are named either Astronomy (37%) or Astrophysics (35%), with the rest split among a variety of other names.

If an institution reported offering multiple Astronomy-related degrees, we asked them to report which degree was more popular and to consider only that degree when responding to the remainder of the survey questions. Among the 27 programs that offer both a BA and a BS, only one program clearly stated that the BA was more popular than the BS option.

There were 18 departments with more than one Astronomy-related degree that provided complete information about both the names and types of majors that they offered. Of these, 11 institutions reported offering a BS in Astrophysics and a BA in Astronomy, specifically. The remainder had a variety of combinations of major names and types.

*2.2.2 Numbers of Majors*

We report here on responses to questions 10 and 14 from the 66 institutions that offer some kind of Astronomy related degree[3]. Where institutions report more than one major, we sum them to report on the entire cohort of students studying Astronomy/Astrophysics at that institution. As invited in the question prompt, some institutions reported a range of cohort size, while others provided a single number estimate. In what follows, we consider the min/max, and an average/median reported number per cohort (calculated as (max-min)/2), noting that where only a single estimate is given these answers are all identical for a given institution. We note here that several institutions also commented on significant variability year–year, which we cannot capture in this analysis.

In this way, we estimate that we received responses from programs that graduate between 718 and 1007 students studying Astronomy/Astrophysics annually. This is comparable to the results from the AIP’s report on Astronomy departments from the 2023-2024 school year[4], in which they reported 982 students receiving an Astronomy-related degree from 92 different US institutions. This same report noted that the number of Astronomy-related degrees awarded in the US has increased by a factor of 200% since 2008-09.

The largest cohort reported in our sample was 75 students, the smallest 0-1, and the average 13.7±14 (1sigma), indicating the wide range of cohort sizes in the sample. It was notable that more than one of the smaller programs commented they are small but currently experiencing growth. E.g. “0 to 1 but growing”, or “we just started, so maybe 1 or 2 per year”

There is no significant difference in cohort size (min, max or average of the average) between institutions offering only one major, versus those offering more than one. We find that Liberal Arts Colleges report having the smallest annual cohorts of astro-related majors (min=0, max=15, average=6.5), followed by Private Research Universities (min=1, max=18, average=7.1) then Public Research Universities (min=1, max=75, average=23). Institutions offering Doctoral or Masters degrees report larger cohorts (min=1, max=75, average=18.4) than those who do not offer higher degrees (min=0, max=20, average=6.5).

*2.2.3 Numbers of required courses*

In this section we combine numerical answers to the total number of courses required for an Astronomy/Astrophysics major from those answering Q11 for single degrees and those answering Q16 for the “most popular degree” out of multiple offerings. Surveyed institutions provided a range of information on the total number of courses offered, from simply ignoring the question, to providing a single number, to a detailed breakdown. Where possible, we converted answers into summary numbers (for a typical class) for:

1. Total number of courses (sum of all physics/astronomy courses + math/CS/chemistry courses/prerequisites);
2. Total number of astronomy courses;
3. Total number of physics + astronomy courses; and
4. Total number of other math/science courses (including math/CS/chemistry etc. required or required as prerequisites).

[3] Three institutions who report offering Astronomy degrees did not provide any information on cohort size.

[4] https://www.aip.org/statistics/roster-of-astronomy-departments-with-enrollment-and-degree-data-2024

We find that on average (for institutions reporting these numbers), 19.4±5 total courses are required for astro-related majors (min 11, max 31), with 5.6±3 (min 1, max 15) of these reported as astronomy/astrophysics courses; 14.8±4 (min 4, max 26) total physics and astronomy courses, and 4.6±2 (min 1, max 12) other required math/science courses (including pre-requisite courses).

Below, we disaggregate these numbers by institution and degree type. We find that BS degrees require slightly more courses (outside of general education/distribution requirements) than BA degrees, and “Astrophysics” slightly more than “Astronomy” (or other names). Public Research Institutions require the largest number of courses, followed by Private Research Institutions, then Liberal Arts Colleges.

**Table 1.** Average numbers of required: total courses (column 2), astronomy-specific courses (column 3), physics+astronomy courses (column 4), and other STEM courses (column 5) for different undergraduate degree types, degree names, and institution types.

| Grouping | Av. Total courses | Astronomy | Physics + Astronomy | Other STEM |
|---|---|---|---|---|
| **All (64)** | **29.4+/5** | **5.6±3** | **14.8±4** | **4.6±2** |
| [redacted] | | | | |
| BA (20) | 15±3 | 4±2 | 11±2 | 3±1 |
| BSc (44) | 21±4 | 6±3 | 17±3 | 5±2 |
| [redacted] | | | | |
| “Astronomy” (16) | 19±5 | 6±3 | 13±4 | 4±2 |
| “Astrophysics” (22) | 20±5 | 7±4 | 16±4 | 5±2 |
| Any other (20) | 19±5 | 4±1.4 | 15±3 | 5±2 |
| [redacted] | | | | |
| Liberal arts college (19) | 16±3.5 | 5±1.6 | 12±3 | 3±1 |
| Private R (15) | 19±4 | 5±1.6 | 15±3 | 5±2 |
| Public R (29) | 21±4 | 6±4 | 17±4 | 5±1.4 |

*2.2.4 Findings*

*2.2.4 Overall Findings on Degree Types and Course Requirements*

***Finding:*** About half of the departments surveyed offer more than one Astronomy-related degree. Of all major titles, “Astronomy” (36%) and “Astrophysics” (30%) were the most common names by far, with the next most common option being a Physics major with a concentration in Astronomy or Astrophysics (15%). Within this, there are some significant correlations between degree name and institution type. Private Research Institutions are more likely to offer “Astrophysics” as the main major compared to other institution types (53% of majors). Public Research Institutions report a greater variety of major names than other institution types.

***Finding:*** The average number of courses required for “Astrophysics” majors was, on average, slightly higher (21±6) than the number of courses required for “Astronomy” majors (19±5), though the difference is not significant.

***Finding***: Of the 18 institutions that offered multiple astro-related majors and reported the names and types of both majors, 11 of these said they specifically offer a BS in Astrophysics and a BA in Astronomy.

***Finding:*** Of the 22 programs that explicitly mention offering both a BS and a BA, 18 of them reported that the BS was the more popular option.

***Finding***: BAs are most common at liberal arts colleges (63%), while make up only 40% of Private Research Institutions and are very rare at Public Research Institutions (of which 93% report BS is most popular)

# 3. Undergraduate Major Course Requirements

All survey respondents who indicated that their department offered one or more Astronomy or Astrophysics degrees were asked the 12 questions below. Respondents from departments offering multiple degrees were asked to answer ***only for their most popular major***.

| No | Question Category | Who Asked | Question Text | Answer Text<br>(Display Logic) |
|---|---|---|---|---|
| 19 | Course Req. | All depts with Astro deg | Which of the following ***mathematics, statistics, and/or computer science*** courses are **required** (including as prerequisites for required physics or astronomy courses) for your undergraduate astronomy/astrophysics major?<br><br>Please also indicate the **number** of required courses by checking the appropriate column<br><br>*Note that **you will specify elective courses that count toward your major later** in this survey. Please indicate here only **required** courses.* | 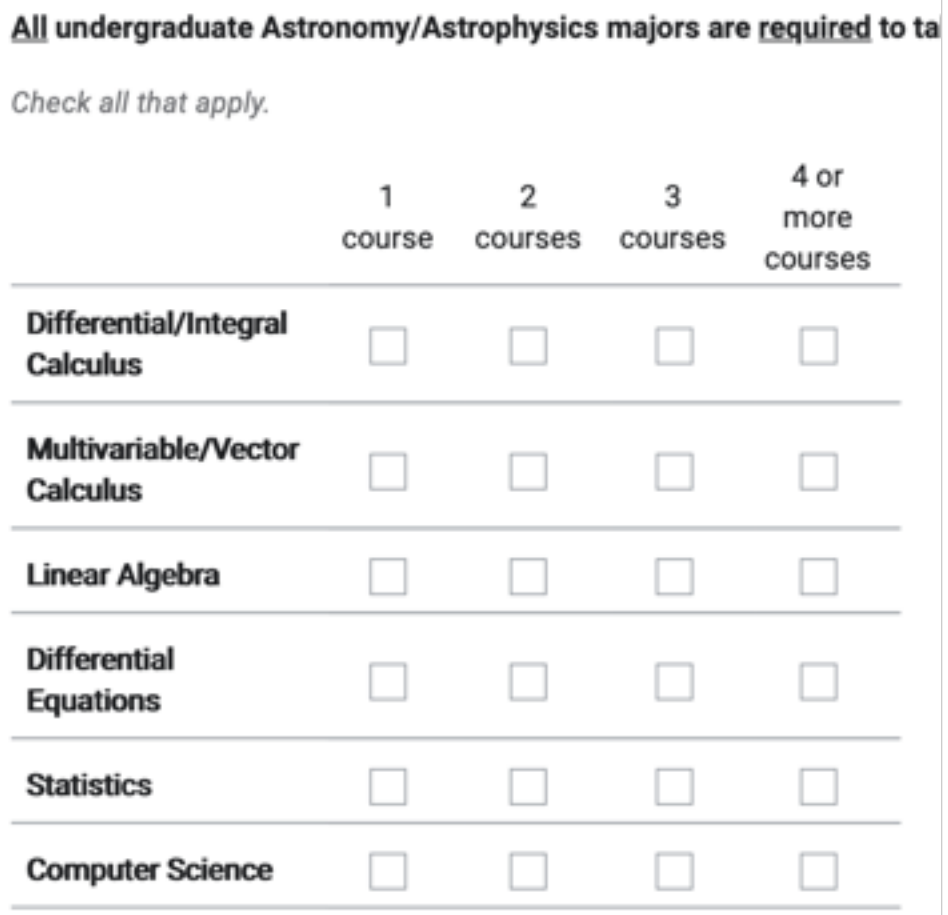<br>All undergraduate Astronomy/Astrophysics majors are required to ta<br>Check all that apply.<br>1 course / 2 courses / 3 courses / 4 or more courses<br>Differential/Integral Calculus<br>Multivariable/Vector Calculus<br>Linear Algebra<br>Differential Equations<br>Statistics<br>Computer Science |
| 20 | Course Req. | All depts with Astro deg | Does your undergraduate major **require** any additional courses in ***mathematics, statistics, or computer science*** not listed above? If so, please list the department, course number, and title of the course below. | Free response |

<table>
<tr><th>No</th><th>Question Category</th><th>Who Asked</th><th>Question Text</th><th>Answer Text<br>(Display Logic)</th></tr>
<tr><td>21</td><td>Course Req.</td><td>All depts with Astro deg</td><td>Which of the following physics courses are <u>required</u> (including as prerequisites) for undergraduate astronomy/astrophysics majors in your department? Please also indicate the level of the course(s) by checking the appropriate column(s).<br><br>For example, if your major requires both a 100 level intro course in electricity and magnetism and a 300 level course in electromagnetic theory, check both the 100 and 300 level boxes in row 2.<br><br>If your institution does not specify course levels with number designations, please translate as best you can</td><td><u>All</u> undergraduate Astronomy/Astrophysics majors are <u>required</u> to tak<br>Check all that apply.
<table>
<tr><th></th><th>100 level</th><th>200 level</th><th>300 level</th><th>400 level or above</th></tr>
<tr><td>Newtonian/Classical Mechanics</td><td>☐</td><td>☐</td><td>☐</td><td>☐</td></tr>
<tr><td>Electricity and Magnetism</td><td>☐</td><td>☐</td><td>☐</td><td>☐</td></tr>
<tr><td>Waves and Optics</td><td>☐</td><td>☐</td><td>☐</td><td>☐</td></tr>
<tr><td>Modern Physics</td><td>☐</td><td>☐</td><td>☐</td><td>☐</td></tr>
<tr><td>Statistical Mechanics and Thermodynamics</td><td>☐</td><td>☐</td><td>☐</td><td>☐</td></tr>
<tr><td>Quantum Physics/Quantum Mechanics</td><td>☐</td><td>☐</td><td>☐</td><td>☐</td></tr>
<tr><td>Computational Physics</td><td>☐</td><td>☐</td><td>☐</td><td>☐</td></tr>
<tr><td>Math Methods for Physicists</td><td>☐</td><td>☐</td><td>☐</td><td>☐</td></tr>
<tr><td>Programming for Physicists</td><td>☐</td><td>☐</td><td>☐</td><td>☐</td></tr>
<tr><td>Experimental or Laboratory Physics</td><td>☐</td><td>☐</td><td>☐</td><td>☐</td></tr>
</table></td></tr>
<tr><td>22</td><td>Course Req.</td><td>All depts with Astro deg</td><td>Does your undergraduate major <u>require</u> any additional courses in physics not listed above? If so, please list the course number and title below.<br><br>You may also use this space to describe any subtleties not captured in the options above (e.g. our 100 level mechanics course also includes one month of waves and optics)</td><td>Free response</td></tr>
<tr><td>23</td><td>Course Req.</td><td>All depts with Astro deg</td><td>How many total astronomy courses are <u>required</u> for your major?<br>please do not count electives in this total</td><td>Radio buttons 1-6 + “other”</td></tr>
</table>

| No | Question Category | Who Asked | Question Text | Answer Text (Display Logic) |
|---|---|---|---|---|
| 24 | Course Req. | All depts with Astro deg | What are your department's **required** ***astronomy*** courses?<br><br>*Please do your best to fit your required courses into the following categories, but also feel free to add additional courses or details in the space provided with the next question*<br><br>***Reminder: you will specify electives in the next section. Please consider only required courses here*** | **All undergraduate Astronomy/Astrophysics majors are required to t**<br>*Check all that apply.*<br>Columns: 100 level / 200 level / 300 level / 400 level or above<br>Rows (checkboxes):<br>Survey course (i.e. "Astro 101", includes nonmajors)<br>Survey course for majors/prospective majors<br>Stars (stellar structure and evolution)<br>ISM<br>Galaxies and Cosmology<br>Galaxies (standalone)<br>Cosmology (standalone)<br>Planetary (solar system AND exoplanets)<br>Solar system (standalone)<br>Exoplanets (standalone)<br>Observational<br>Programming or data science for Astronomy<br>Research capstone (1 semester)<br>Research capstone (2 semester) |
| 25 | Course Req. | All depts with Astro deg | Does your undergraduate major **require** any additional courses in ***astronomy*** not listed above? If so, please list the course number and title below.<br><br>*You may also use this space to describe any subtleties not captured in the options above (e.g. our stars course also includes 1 week of ISM).* | Free response |
| 26 | Course Req. | All depts with Astro deg | How many ***astronomy electives*** are required to complete your department's astronomy/astrophysics major? | Radio buttons 1-5 + "other" |
| 27 | Course Req. | All depts with Astro deg | Which of the following are offered as options for ***astronomy electives*** toward your astronomy/astrophysics major? | Same as Q24, but with the preamble "**Astronomy/Astrophysics majors are allowed to count as electives toward the major:**" |

| No | Question Category | Who Asked | Question Text | Answer Text (Display Logic) |
|---|---|---|---|---|
| 28 | Course Req. | All depts with Astro deg | Does your undergraduate major have any additional options for elective courses in ***astronomy*** not listed above? If so, please list the course number and title below. | Free response |
| 29 | Course Req. | All depts with Astro deg | How many **total electives** (in all disciplines, including Astronomy) are **required** for your major? *note: we are asking about electives that count directly toward the major, NOT distribution or general education requirements* | Radio buttons 1-5 + "other" |
| 30 | Course Req. | All depts with Astro deg | What are the requirements and options for other disciplines' courses serving as **electives** toward your astronomy/astrophysics major? *e.g. two physics courses at the 300 level or higher, or "three 200+ level statistics or computer science classes"* | Free response |

## *3.1 Summary and Key Takeaways*

There is ***much*** variety in required coursework among programs offering majors in Astronomy and Astrophysics. In fact, the ***only*** coursework required by ***all*** programs is differential and integral calculus, lower-division classical mechanics, and lower-division electricity and magnetism, mirroring the findings of Cabanela and Partridge (2002). Almost all programs require multivariable calculus and at least one course in differential equations. Among Astronomy courses specifically, the most popular requirements are a lower-division survey course for majors (required by 49.2% of programs) and upper-division courses on stars (43.1%), observational astronomy (40.0%), and galaxies and cosmology (30.8%).

## *3.2 Data Summary and Analysis*

We received 88 total survey responses representing 78 unique institutions. Of those institutions, 12 do not offer a major in Astronomy, so they are excluded from the analyses in this section. Out of the remaining 66, nine were represented by multiple responses to the survey. For these institutions, we combined the information from all respondents to create a single entry for that institution. Whenever there was a disagreement between respondents, we consulted the degree requirements listed on that institution's website and updated that institution's entry in our data set accordingly.

We also read the answers to the free-response questions (20, 22, 25, 28, and 30). Occasionally, they mentioned a class that should have been marked in one of the multi-select questions but was not (for whatever reason). We updated the entries for these institutions to include these classes. We also double-checked the responses for several institutions against the information available online about the course requirements for their majors. This was done when we encountered entries that seemed suspicious (e.g., omitting differential calculus from reported required classes). Again, we updated that institution's entry in our data set when appropriate based on the information on their website.

A number of institutions required multiple courses in a specific topic at the same level (E&M I and II was the most common). The survey only allowed for a checkbox selecting the levels of the required courses, so there was no way for respondents to report multiple courses at the same level.

Websites also reveal that several institutions require Astronomy courses that don't fit neatly into the provided categories and also weren't mentioned in free responses. We checked responses to the survey against posted course requirements wherever possible, and any discrepancies are indicated in the "Notes" column (Column MD) of the "Course Requirement Analysis" tab of the public dataset.

Figure 6 shows the percentage of programs that require coursework in statistics, computer science, and various specific math disciplines (differential and integral calculus, multivariable calculus, linear algebra, and differential equations). Since programs often use different numbers of courses to cover similar material, we report our results as simply a binary "coursework required" in a particular subject, rather than the number of required courses. All programs, unsurprisingly, require coursework in differential and integral calculus, however there is a divergence after introductory calculus. Most (92.3%), but not all, require coursework in multivariable/vector calculus. Many also require coursework in linear algebra (60.0%) and/or differential equations (75.4%). Very few (7.7%) require statistics. About a third, 30.8%, require coursework in computer science. However, we note that there are many programs that offer their own programming and/or data science courses specific to physics and/or astronomy, which may be why they do not require separate computer science courses.

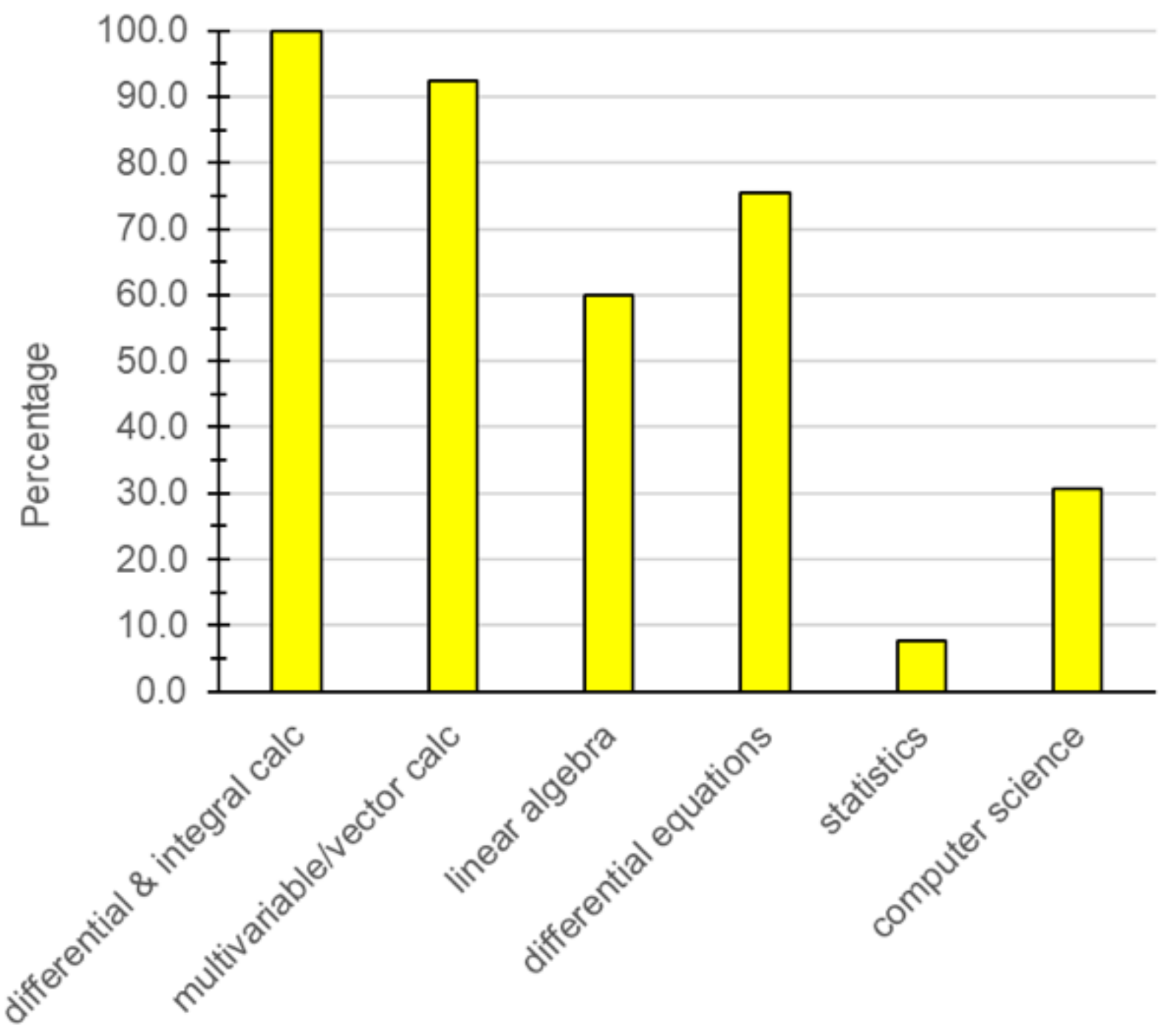

**Figure 6.** Percentage of reporting degree programs that require coursework in various mathematical topics, statistics, and computer science.

The top of Figure 7 shows the percentage of programs requiring lower-division coursework in various physics subjects. The bottom of Figure 7 shows the same information, but for required upper-division coursework. Although we asked respondents to list whether a given course was at the 100-, 200-, 300-, or 400-level or above, different institutions use different numbering systems, so it is difficult to report and analyze the data at that level of granularity. For example, some institutions use 100-level to represent "freshman" level courses. Others use numbers in the 100s to represent courses for non-majors and numbers in the 200s to label their introductory physics courses for majors. We decided to designate all 100- and 200-level courses as "lower-division," since they are typically taken by students in their first two years of college. Since 300- and 400-level courses are generally taken during students' junior and senior years, we refer to these courses as "upper-division." Figures 7 and 8 show that lower-division classical mechanics and electricity and magnetism are the ***only*** physics courses required by all Astronomy/Astrophysics undergraduate programs.

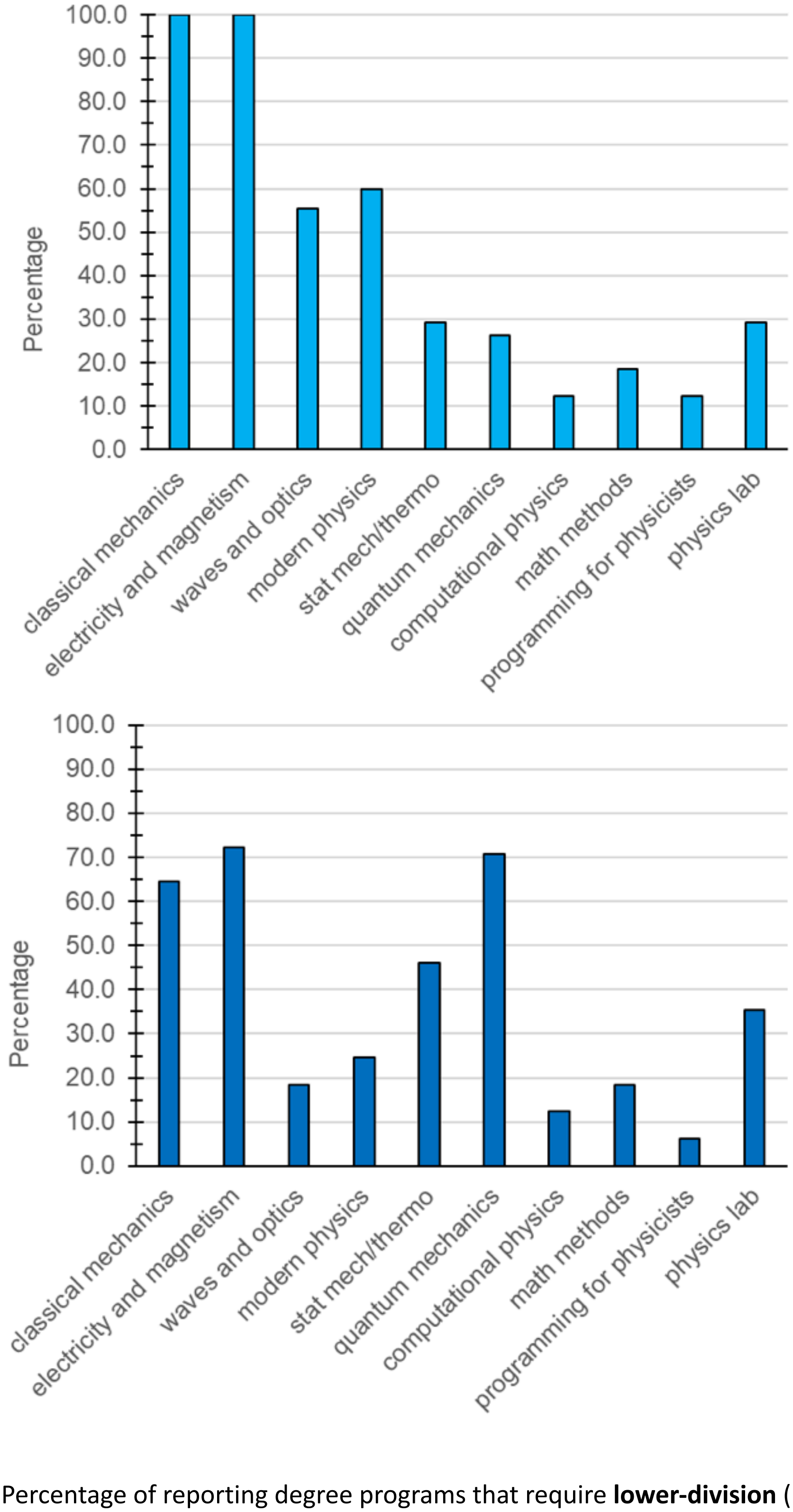

**Figure 7.** *Top*: Percentage of reporting degree programs that require **lower-division** (100-200 level) physics coursework in the listed subjects. *Bottom*:Percentage of reporting degree programs that require **upper-division** (300-400 level) physics coursework in the listed subjects.

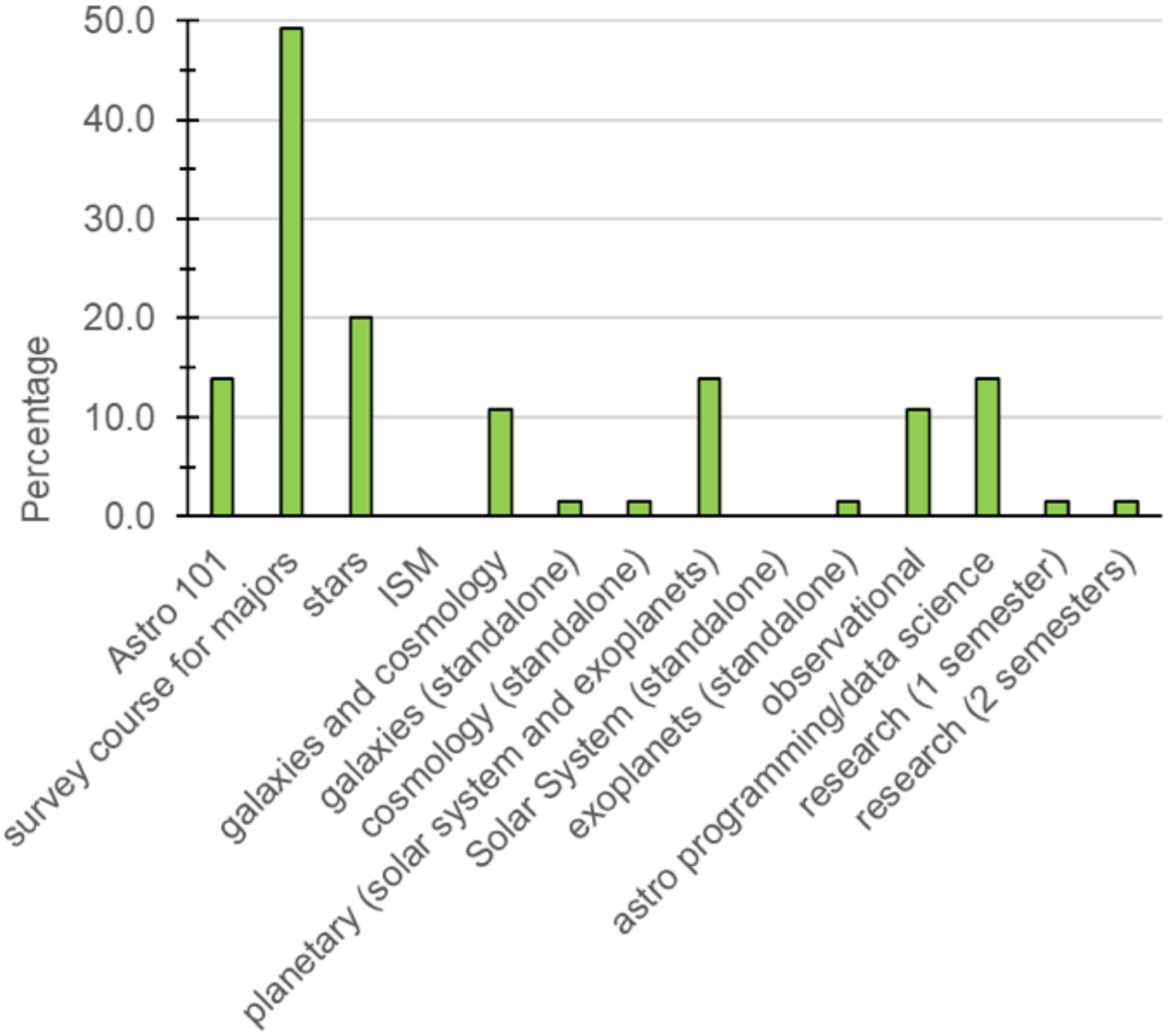

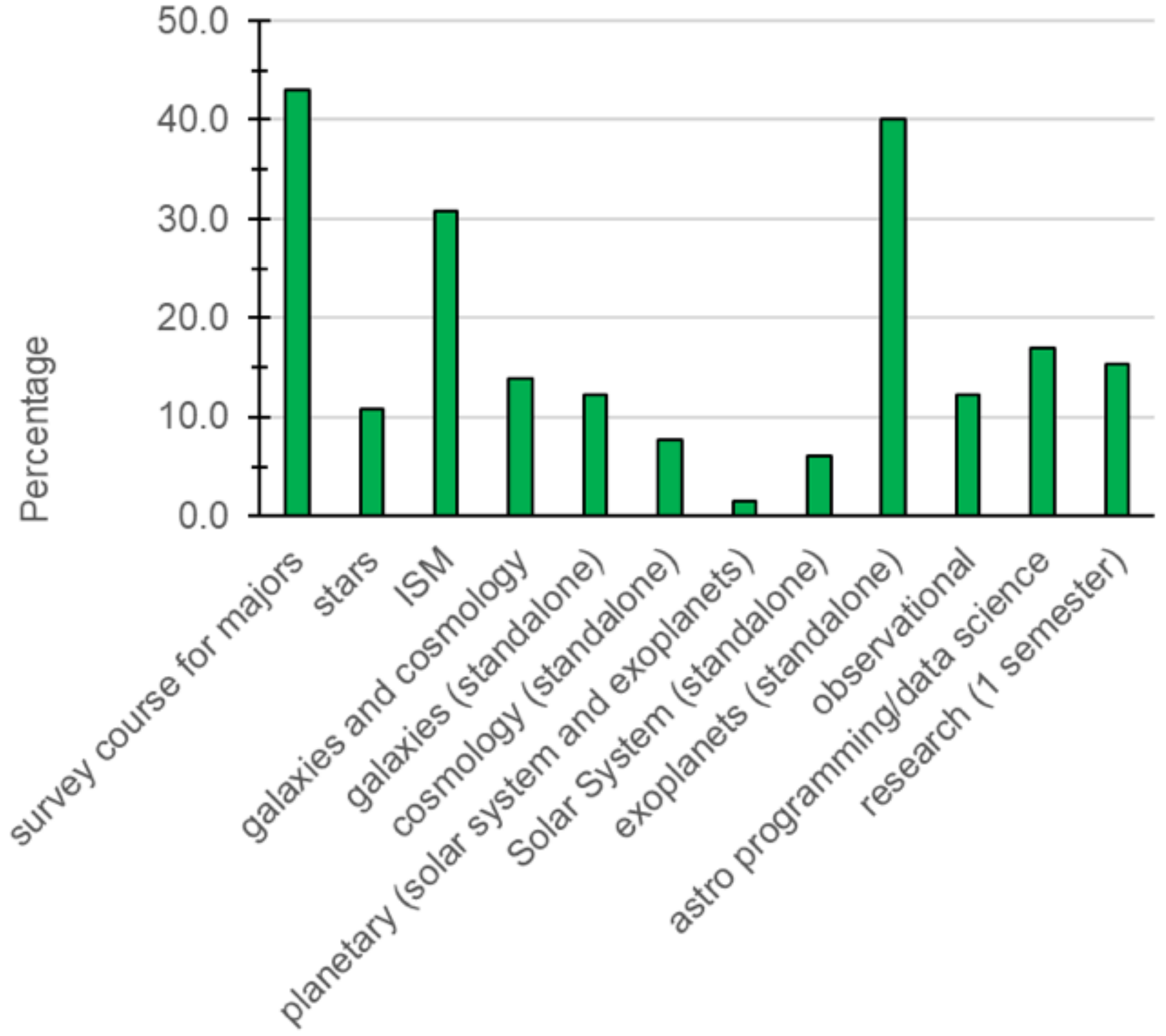

**Figure 8.** *Top:* Percentage of reporting degree programs that require **lower-division** (100-200 level) astronomy coursework in the listed subjects. *Bottom:* Percentage of reporting degree programs that require **upper-division** (300-400 level) astronomy coursework in the listed subjects.

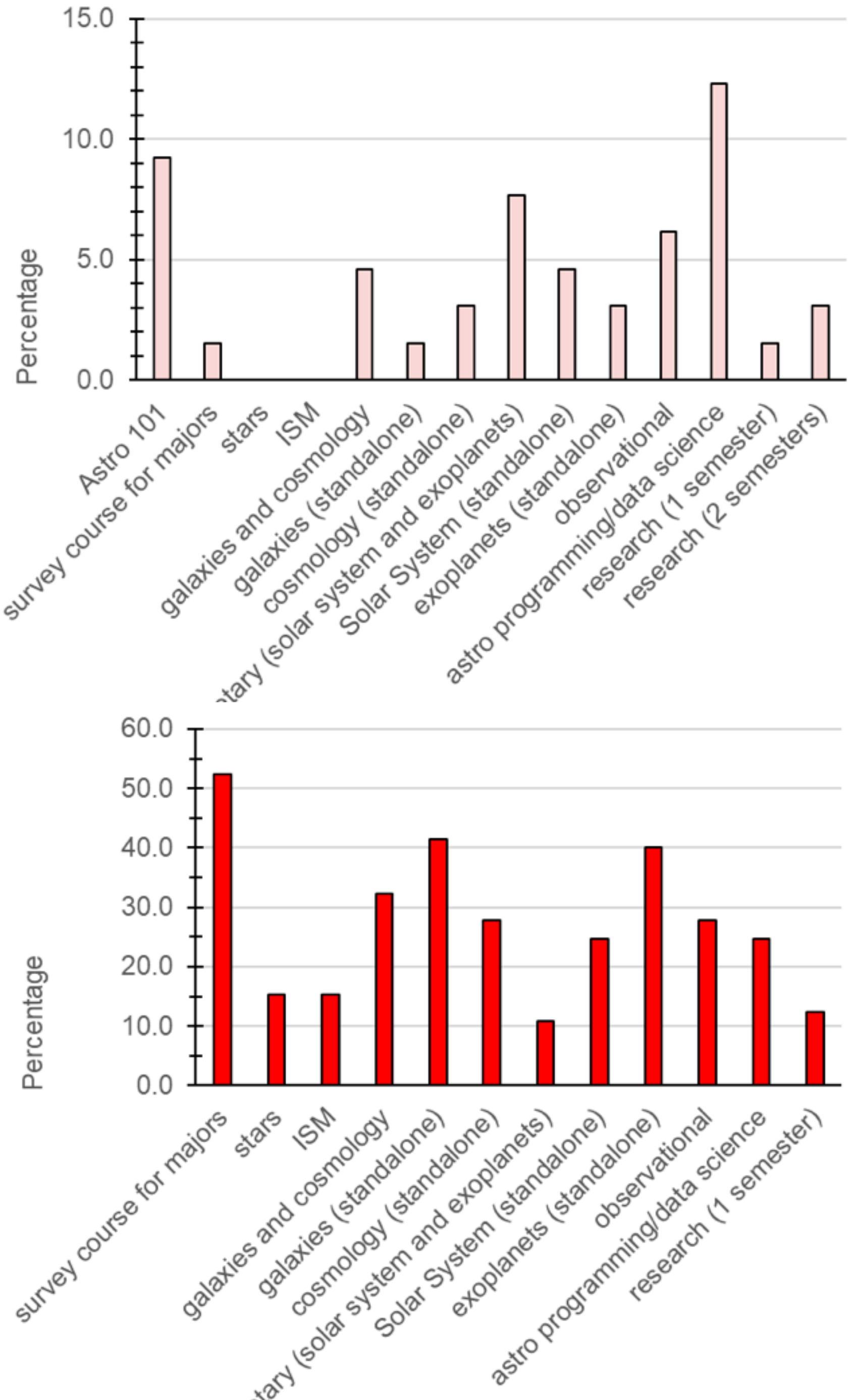

**Figure 9.** *Top:* Percentage of reporting degree programs that offer **lower-division** (100-200 level) astronomy electives in the listed subjects. *Bottom:* Percentage of reporting degree programs that offer **upper-division** (300-400 level) astronomy electives in the listed subjects.

Figure 8 shows the percentage of Astronomy programs **requiring** coursework in the listed subjects, at both the lower- and upper-division levels on top and bottom, respectively. Note that "Astro 101" is an umbrella term for any introductory astronomy course that is also taken by non-majors. While some programs require this course for their majors, others have a separate survey course that is only for students majoring in Astronomy and related disciplines.

Figure 9 shows the percentage of Astronomy programs that offer **elective** coursework in the listed subjects at the lower- and upper-division levels, shown on top and bottom, respectively. We note that some programs offer classes that are not among the subjects that were listed in our survey. These most commonly include classes on radiative processes, general relativity, and high-energy astrophysics.

# 4. Undergraduate Major Learning Goals

All respondents from departments offering undergraduate degrees in Astronomy and/or Astrophysics were asked questions 31-32 below. Those responding yes to question 32 were asked two additional questions (Q33-34)

| No | Question Category | Who Asked | Question Text | Answer Text (Display Logic) |
|---|---|---|---|---|
| 31 | Goals | All depts with Astro deg | What are the **top three things** you think every major from your program should know or be able to do upon completing your program? | Free response |
| 32 | Goals | All depts with Astro deg | Does your department have **formal learning goals or objectives** for your undergraduate degree? | Yes<br>No → Q35<br>I'm not sure → Q35 |
| 33 | Goals | All depts with learning goals | What are the stated learning goals or objectives of your undergraduate major(s) in astronomy and/or astrophysics? | Free response |
| 34 | Goals | All depts with learning goals | How and how often, if at all, does your department assess whether it has met these learning goals/objectives? | Free response |

## *4.1 Summary and Key Takeaways*

63 survey respondents provided responses to Question 31, which read: "What are the top three things you think every major from your program should know or be able to do upon completing your program?". We consider these responses to be an *informal* measure of individuals' views regarding the learning goals of an Astronomy major, and our aim was to contrast the nature and frequency of responses to this question to the *formal* learning goals respondents provided in Question 33.

Individuals from 76 unique departments answered one or more of Questions 32-34 about formal learning objectives, with 45 (59%) reporting that their department ***does*** have formal learning objectives, 23 (30%) reporting that their department ***does not*** have formal learning objectives, and 8 (11%) reporting "I'm not sure" (4 current/former students, 4 faculty).

In analyzing responses to both questions 31 and 33, a committee of 3 SURGE members first examined the responses in bulk and inductively identified 14 common themes, which are listed in the first column of Table 2. Each of the 3 members then individually coded the survey responses by flagging themes met by each write-in response. In the final analysis, a survey response was only classified as meeting a given theme if 2 or more of the coders flagged it as such.

**Table 2.** Response rates and relative rankings for each learning goal category among formal (column 2) and informal (column 3) learning goals. The color and shade of the final column indicates the *relative* priority of a given learning goal between informal and formal learning goals, with green indicating a higher and pink indicating a lower priority ranking among informal learning goals compared to formal.

| Category | **Formal Learning Goals**<br>% of Responses (n=45)<br>***Rank*** | **Informal Learning Goals**<br>% of Responses (n=62)<br>***Rank*** | **Difference** |
|---|---|---|---|
| Physics Knowledge and Skills | 64%<br>***#1*** | 31%<br>***#6*** | -33%<br>***-5 ranks*** |
| Problem Solving or Critical Thinking Skills | 56%<br>***#2*** | 32%<br>***#4*** | -24%<br>***-2 ranks*** |
| Communication Skills | 53%<br>***#3*** | 34%<br>***#3*** | -19%<br>***same rank*** |
| Observational, Experimental, Instrumentation or Laboratory Skills | 51%<br>***#4*** | 29%<br>***#7*** | -22%<br>***-3 ranks*** |
| Astronomy and Astrophysics Knowledge and Skills | 47%<br>***#5*** | 60%<br>***#2*** | +13%<br>***+3 ranks*** |
| Data, Computational, Programming Skills | 47%<br>***#5*** | 63%<br>***#1*** | +16%<br>***+4 ranks*** |
| Mathematics, Statistics, or Quantitative Knowledge and Skills | 44%<br>***#7*** | 32%<br>***#4*** | -12%<br>***+3 ranks*** |
| Scientific Reasoning, Method, or Process Skills | 29%<br>***#8*** | 23%<br>***#9*** | -6%<br>***-1 rank*** |
| Research Experience and Skills | 29%<br>***#8*** | 26%<br>***#8*** | -3%<br>***same rank*** |
| Ethical and Inclusion Training or Skills | 22%<br>***#10*** | 0%<br>***#12*** | -22%<br>***-2 ranks*** |
| Training for industry, careers, or the profession broadly | 20%<br>***#11*** | 0%<br>***#12*** | -20%<br>***-1 rank*** |
| Collaboration Skills | 15%<br>***#12*** | 0%<br>***#12*** | ***-15%***<br>***same rank*** |
| Preparation for graduate work | 13%<br>***#13*** | 5%<br>***#11*** | -8%<br>***+2 rank*** |
| Reading Academic Literature or Finding Scientific Information | 11%<br>***#14*** | 6%<br>***#10*** | -5%<br>***+4 ranks*** |

Table 2 summarizes the results of our analysis regarding the frequency with which each of the themes was identified among formal (column 2) and informal (column 3) learning goal survey responses. Whereas Question 31 asked participants to list just *three* things individuals should take from an Astronomy/Astrophysics major, many departments' formal learning goals number many more than three. In this way, the responses to questions 31 and 33 are not an “apples to apples” comparison, and respondents to question 31 are asked to engage in a prioritization that is not present in Question 33.

There is, unsurprisingly, a greater dispersion among answers to Question 31, as evidenced by a decrease in the overall proportion of learning goal responses meeting each of the 13 themes (with 2 exceptions, discussed in detail below). For this reason, we also provide in Table 2 the relative *ranking* of each learning goal theme among the responses, in addition to the proportion, which we assert is a more direct comparison. The final column lists both the difference in proportion of responses meeting each theme and the difference in the ranking of that theme between formal and informal learning goals.

There are several differences worth highlighting among these relative rankings. The first is the two exceptions to the general rule described above that responses to the informal learning goals were less aligned (as indicated by a lower overall proportion of respondents specifying that theme) than formal learning goals. Two themes, namely “Data, Computational, and Programming Skills” and “Astronomy and Astrophysics Knowledge and Skills” were mentioned in more than 60% of informal learning goal responses, placing them in the 1 and 2 ranks, respectively. Yet, among formal learning goals, these two themes are tied at Rank 5. This indicates, perhaps, a mismatch between stated learning goals and the skills that we value most highly among Astronomy and Astrophysics graduates, though we note that neither theme is by any means absent among formal learning goals, with nearly half of departments listing each of them among those goals.

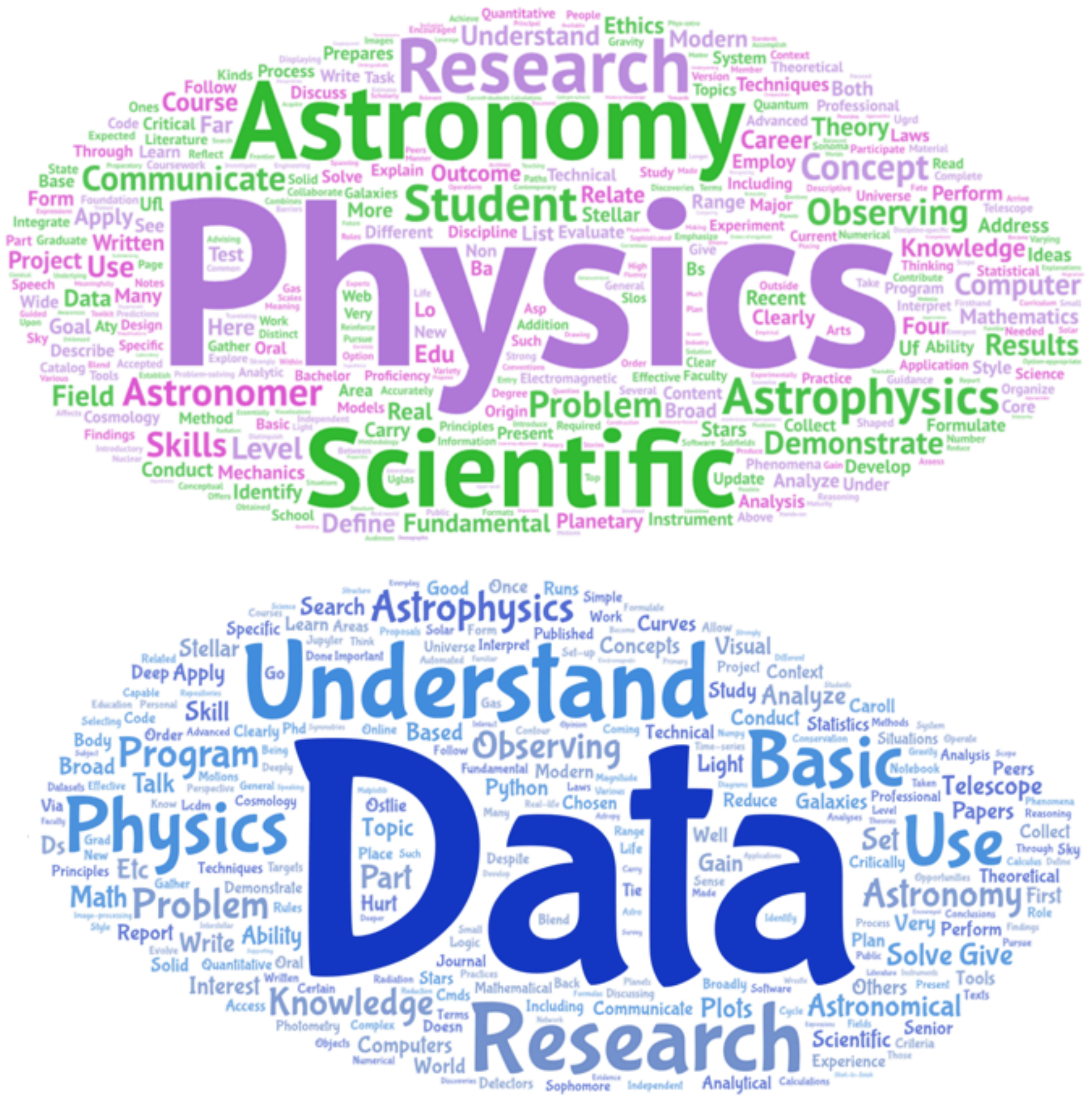

**Figure 10.** Word clouds derived from the raw text responses to Questions 33 (top, formal learning goals) and 31 (bottom, informal learning goals), where size indicates the frequency with which that term appears among responses.

## 4.2 Data Summary and Analysis

### 4.2.1 Learning Goals

Tables 3 and 4 present the results of our thematic analysis of informal and formal learning goals, respectively.

**Table 3.** Identified themes (column 1) among Informal Learning goal responses (Question 31), with total number of responses meeting that theme (of 63 total, column 2) and examples (column 3).

| Theme | No. | Sample Responses |
|---|---|---|
| Data, Computational, Programming Skills | 39 | ● Wrestle with data! Collect a data set, interact with it via programming, analyze it using basic statistics<br>● find, explore, analyze, and visualize data (astronomical and otherwise) |
| Astronomy and Astrophysics Knowledge and Skills | 37 | ● Have a broad understanding of our knowledge of modern astrophysics and the evidence supporting those discoveries<br>● Identify basic concepts from many areas of astronomy, including motions in the sky, gravity, electromagnetic radiation, solar system, stars and galaxies |
| Communication Skills | 21 | ● write effectively about astrophysics and science more generally<br>● communicate in oral and written form scientific ideas |
| Problem Solving or Critical Thinking Skills | 20 | ● approach/frame and solve complex problems on size scales that range from nuclear to cosmological<br>● Prepared to use mathematical tools and original computer code to solve real physical problems |
| Math, Stats, or Quantitative Knowledge and Skills | 20 | ● formulate scientific problems in mathematical terms<br>● Solve quantitative problems |
| Physics Knowledge and Skills | 19 | ● Being able to describe the physical behavior of astronomical objects<br>● apply physics thinking in new situations |
| Observational, Experimental, Instrumentation or Laboratory Skills | 18 | ● Plan observing runs on astronomical objects by selecting targets based on certain criteria, perform real-life observations using a modern telescope set-up (automated or not), use image-processing software to reduce data, and perform basic photometry on data to generate CMDs, time-series light curves, and such.<br>● Know how astronomical data is taken, processed, and analyzed so data isn't a black box to them. |
| Research Experience and Skills | 16 | ● Develop the skills to carry out a small independent research project. Learn to define the scope of the project, conduct an effective literature search, perform computations, and analyze data.<br>● Effectively conduct research through the use of programming, software, mathematical techniques, and concepts in astrophysics. |
| Scientific Reasoning, Method, or Process Skills | 14 | ● understand the scientific process at a very deep level<br>● understand the role of scientific thinking in the everyday world |
| Reading Academic Literature or Finding Scientific Information | 4 | ● How to read a journal article, interpret what the authors did, and explain that to audiences ranging from expert Astronomers to non-scientists.<br>● Study published professional journal papers and either write about them or give a talk about them |

**Table 4.** Identified themes (column 1) among Informal Learning goal responses (Question 33), number of responses meeting that theme (of 45 total, column 2) and sample responses (column 3).

| Theme | No. | Sample Responses |
|---|---|---|
| Physics Knowledge and Skills | 29 | ● Conceptual and analytical understanding of the four major areas of physics: Mechanics, Electricity & Magnetism, Quantum Mechanics, and Statistical Physics<br>● Students will solve (using the appropriate mathematical techniques) any advanced undergraduate problem from the core areas of physics (Newtonian mechanics, electromagnetism, statistical mechanics, quantum theory, and relativity) |
| Problem Solving or Critical Thinking Skills | 25 | ● Be skilled at quantitative problem solving incorporating hypothesis formation, data analysis, error analysis, conceptual modeling, numerical computation, and hypothesis testing through quantitative comparison between observation and theoretical concepts.<br>● Identify, formulate, and solve tractable scientific and technical problems by placing them in context, making appropriate estimates and simplifications, modeling the important physical processes, quantifying predictions with analytic and computational tools, and testing the correctness of the results. |
| Communication Skills | 24 | ● Effectively and clearly communicate ideas and results in speech and in writing in an accepted style of presentation.<br>● Develop proficiency with communicating, translating and interpreting fundamental astronomical concepts or research results in oral and written formats. |
| Observational, Experimental, Instrumentation or Laboratory Skills | 23 | ● Be familiar with scientific instrumentation used by professional astronomers. Be familiar with digital imaging as a source of scientific data, including techniques of acquisition, reduction, and analysis.<br>● Use telescopes, instruments, and computers to gather and reduce astronomical data. |
| Astronomy and Astrophysics Knowledge and Skills | 21 | ● Knowledge of the contents of the extraterrestrial universe, including planets, stars, galaxies, and the large-scale structure of the universe itself, and understanding of the formation and evolution of all of these.<br>● apply astronomical concepts and principles of physical science to articulate, discuss and explicate core disciplinary concepts in astronomy and planetary sciences. |
| Data, Computational, Programming Skills | 21 | ● Write original computer code to accomplish a computational task, such as analyzing data, displaying astronomical images, or performing calculations.<br>● manage, evaluate, and interpret datasets to solve quantitative problems and test hypotheses within an astronomical context utilizing modern computing methods |
| Math, Stats, or Quantitative Knowledge and Skills | 20 | ● Examine a physical situation, construct a quantitative model of the situation, analyze the model analytically or numerically, and evaluate the accuracy of the model.<br>● Be skilled at quantitative problem solving incorporating hypothesis formation, data analysis, error analysis, conceptual modeling, numerical computation, and hypothesis testing through quantitative comparison between observation and theoretical concepts. |
| Scientific Reasoning, Method, or Process Skills | 13 | ● Examine a physical situation, construct a quantitative model of the situation, analyze the model analytically or numerically, and evaluate the accuracy of the model.<br>● Students will analyze which physical processes are relevant to a given system. Students will assess cause and effect in physical systems by formulating evidence-based logical arguments. |

| Theme | No. | Sample Responses |
|---|---|---|
| Research Experience and Skills | 13 | ● firsthand experience with the process of science through participation in research<br>● Develop the skills to carry out a small independent research project. Learn to define the scope of the project, conduct an effective literature search, perform computations, and analyze data. |
| Scientific Ethics or Inclusion Skills or Knowledge | 10 | ● Recognize the role that historic and social factors play in the practice of physics or astronomy, as well as the ways physics or astronomy has influenced history and society.<br>● understand scientific ethical practices and demonstrate them in the conduct of scientific research. |
| Industry, Career, or Professional Skills or Knowledge | 9 | ● Generate fluency in the scientific enterprise and awareness of possible career paths available to the undergraduate astronomy and astrophysics major.<br>● Gain entry to top graduate programs, employment as physicists in industry, teaching positions in high school physics and astronomy, or leverage their skills in other rewarding careers |
| Collaboration Skills | 7 | ● demonstrate the ability to work collaboratively in interdisciplinary groups to solve scientific problems<br>● Collaborate with peers on research projects that address scientific and technical problems using experiments, computer models, and analysis. |
| Preparation for Graduate Work | 6 | ● technical and research skills needed to pursue graduate study in astronomy<br>● The program provides balanced and integrated coursework in astronomy, mathematics, and physics that prepares students for graduate studies in astronomy, astrophysics, or related science disciplines |
| Reading Academic Literature or Finding Scientific Information | 5 | ● Learn how to read and critically evaluate scientific literature<br>● analyze and evaluate scientific information in order to describe a question at the frontier of an astronomical discipline. |

#### *4.2.2 Frequency and Methods of Learning Goal Assessment*

Of the 45 departments that listed formal learning goals, 31 departments indicated the frequency at which they assess whether these learning goals are being met. From most often to least often, these are: 2 departments (6%) more than once annually, 16 departments (52%) annually, 5 departments (16%) every 2-4 years, 7 departments (23%) every 5-10 years, and 1 department “irregularly”. An additional 4 departmental respondents were unsure about the frequency with which their departments evaluated progress toward learning goals, and 7 indicated that their departments did not formally evaluate whether their learning goals were being met.

14 departments also provided enough information about ***how*** their department evaluated whether learning goals were being met. These responses can be categorized broadly into (a) formal external reviews (n=6) and (b) informal department- or course- (especially capstone-course) level reviews (n=9). We believe that evaluating, formally or informally, departmental efficacy in meeting learning goals or objectives is a best practice. We thus provide the text of these responses in Table 5 so that they might serve as examples for other departments looking to evaluate the efficacy of or their learning goals.

**Table 5.** The text of all 14 survey responses to question 34 where respondents indicated how their departments review progress toward meeting their adopted learning goals, separated into formal reviews (column 1) and informal reviews (column 2)

| External formal reviews | Department-level reviews |
|---|---|
| ● data is collected every year and most formally assessed during academic program reviews every ~ 7 years<br>● Yearly through accreditation reports<br>● Programmatically and formally, every 5 years. The Major was introduced in 2018, so we have only undergone one program assessment at this point.<br>● We are formally reviewed on a 3 year timescale, but since we are newly rolling out our program, we are continuously assessing and making tweaks<br>● There is a yearly review that is expected from the University level. We must report on how we met the goals, identified strengths in the program, challenges that need addressed, and how we plan to meet those challenges. There is a wide range of assessment tools that go into this process that would be hard to cover in this survey.<br>● Our program is formally reviewed every 5 years. | ● At advising meetings for registering for classes, especially so in the latter half of the undergraduate career when faculty advisers are assigned.<br>● aside from specific course content, we use the capstone projects to assess graduating students in these areas.<br>● More frequently and informally, we review how individual courses are meeting learning goals and work to keep them aligned across the curriculum.<br>● Projects submitted as part of the Observational Astronomy course and the advanced Stellar/Galactic courses<br>● Exams, Portfolios, Papers, Oral presentations<br>● we discuss the capstones each year<br>● ~ once yearly in Astronomy Dept. meeting<br>● We review learning goals annually in department meeting.<br>● The outcomes of our program are regularly evaluated through several channels. Students provide course feedback to faculty and teaching assistants directly and through the online TQFR system; graduating students fill out an exit survey; the astrophysics option representative meets with students regularly; the students in the program collectively discuss the program annually with representatives of the faculty as part of the Student Faculty Conference. In addition, alumni outcomes are monitored at annual national astronomy meetings attended by a large fraction of our alumni. The information gathered is discussed in faculty meetings and used to improve class teaching and professor assignments, and to motivate curriculum changes. |

# 5. Further Departmental Information

Survey respondents were asked about various activities their Astronomy majors might participate in besides coursework, such as research, a thesis, a second major, and attending graduate school (Q35). The responses to each question are summarized in 5.1, and a synthesis is presented in 5.2.

| No | Question Category | Who Asked | Question Text | Answer Text (Display Logic) |
|---|---|---|---|---|
| 35 | Other | All depts with Astro deg | To the best of your knowledge, approximately what proportion of your majors do each of the following? | 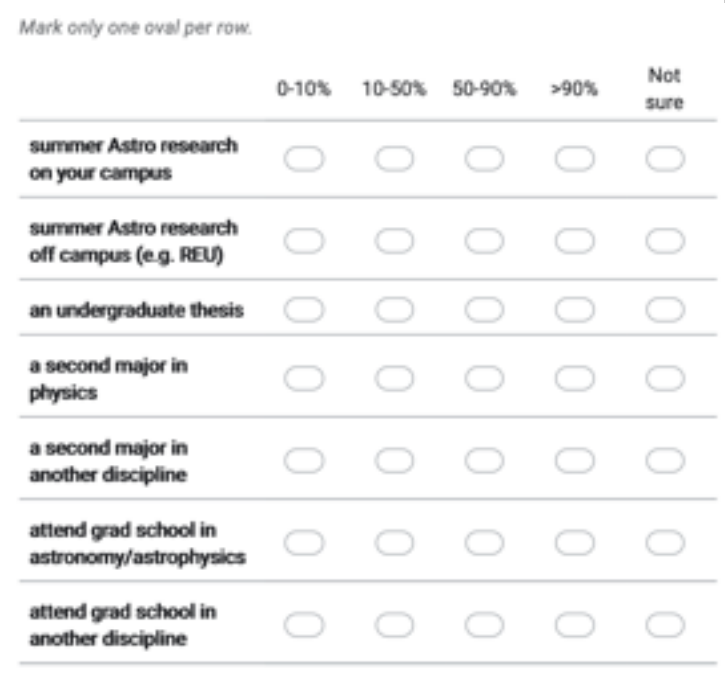<br>*Mark only one oval per row.*<br>0-10% 10-50% 50-90% >90% Not sure<br>summer Astro research on your campus<br>summer Astro research off campus (e.g. REU)<br>an undergraduate thesis<br>a second major in physics<br>a second major in another discipline<br>attend grad school in astronomy/astrophysics<br>attend grad school in another discipline |
| 36 | Other | All depts with Astro deg | Does your department have a comprehensive or capstone requirement for graduating students? If so, please describe it below.<br><br>*e.g. students must attend 9 departmental seminars and present a paper to a panel of 3 faculty* | Free response |
| 37 | Other | All depts with Astro deg | Approximately what proportion of students who declare a major in Astronomy or Astrophysics in your department complete it? What, in your opinion, drives away those who don't?<br><br>*We ask this question only to identify patterns in responses that might inform our work as a committee. Your responses to this question will ONLY be reported in aggregate and will not be associated with your department in our report* | Free response |
| 38 | Other | All depts with Astro deg | Is there anything else that you think we should know about your department's undergraduate curriculum? For example: best practices, outcomes, needs, frustrations, or patterns over time?<br><br>*We ask this question only to identify patterns in responses that might inform our work as a committee. Your responses to this question will ONLY be reported in aggregate and will not be associated with your department in our report* | Free response |

## *5.1 Summary and Key Takeaways*

The responses to questions 35–38, taken together, reveal a national landscape in which undergraduate Astronomy and Astrophysics programs share broadly similar goals (fostering research engagement, scientific literacy, and student persistence) but vary considerably in the resources, structures, and challenges shaping how those goals are achieved. Departments of all sizes emphasize the importance of undergraduate research, whether through required capstone experiences, mentored independent studies, or integration of research elements into coursework. Yet while research is nearly universally valued, many institutions report difficulty offering sufficient opportunities, citing limited faculty capacity, oversubscription of national REU programs, and rising student interest that outpaces available mentorship.

Patterns in student progression and completion echo these capacity issues. Programs with close faculty–student interaction and cohesive advising communities report high retention, while those constrained by large enrollments or limited staffing note attrition driven by early struggles in math and physics, misconceptions about the major's quantitative rigor, and anxiety about future career paths. The most successful programs appear to mitigate these barriers through early research exposure, structured mentoring, and explicit attention to student belonging and confidence in technical skills.

In curricula, departments are actively experimenting with how best to balance rigor, flexibility, and accessibility. Many are revising degree requirements to accommodate double majors, better sequence programming and data-analysis instruction, and integrate writing or communication skills. Several report that general education requirements or institutional policies constrain their ability to add new courses, while others highlight innovative interdisciplinary designs linking astronomy to engineering, geoscience, or computer science. Across the board, respondents describe a growing need for curricular coherence and advising support as enrollments rise and student preparation becomes more uneven.

Finally, the comments point to a shared awareness that the undergraduate experience in astronomy sits at the intersection of opportunity and strain. Faculty take pride in the engagement and enthusiasm of their students, yet express concern about sustainability, both in mentoring load and in systemic pressures such as the expectation of publication before graduate school. Departments are navigating these tensions with creativity and commitment, experimenting with new instructional models and community structures to ensure that undergraduates not only complete their degrees, but also develop as capable, confident scientific thinkers.

## *5.2 Data Summary and Analysis*

Responses to Question 35 are summarized graphically in Figure 11. On each question, there is a wide range of variability in the data: a significant number of programs, for example, gave each of the possible answers for 35c (*"what proportion of your majors do… an undergraduate thesis?*"), reflecting the diversity among what is typical in undergraduate Astronomy programs. Some trends are present: the response distribution for sub-questions (a), (b), (e), (f), and (g) were similar (generally: a mode of "10-50%" which contained a large plurality of responses; a runner-up of either "50-90%" or "0-10%"; and few responses ">90%"), indicating that **for most programs**, there is a **wide diversity among student engagement** in **research experiences** and **graduate school plans**. In other words, most programs contain a highly heterogeneous mix of student experiences; it is rare for a department to consist of students with highly similar paths.

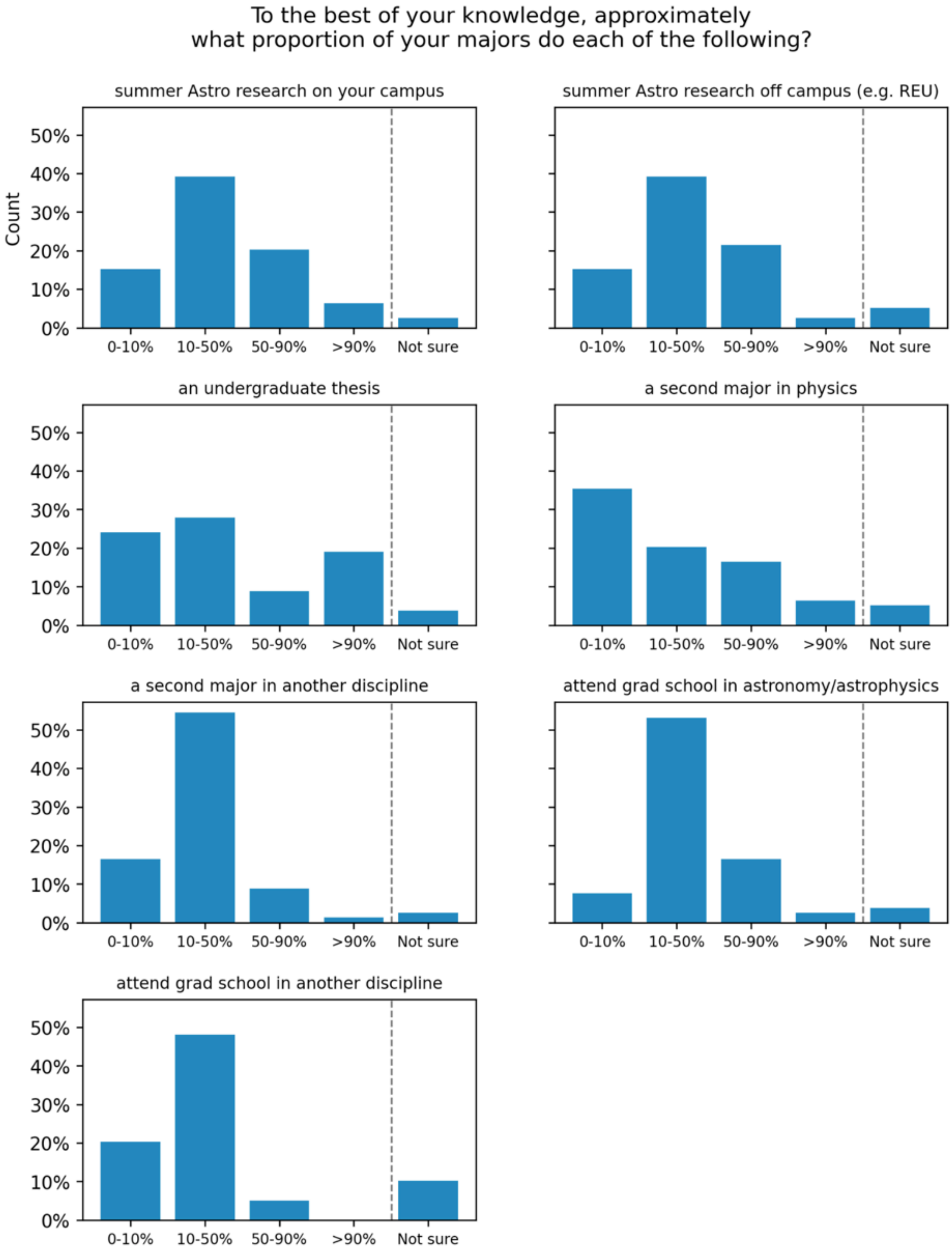

**Figure 11.** Histograms of responses to survey Question 35, indicating the rates at which students in various programs participate in various non-coursework pursuits.

Responses to Question 36, which asked about comprehensive or capstone requirements, largely fell into three categories, namely:

- **Required** capstone/thesis (approx. 40%)
- **Optional** capstone or thesis (approx. 35%)
- **No** formal capstone requirement (approx. 25%)

"Capstones" were most commonly described as a research project or thesis presented either as a written report, an oral/poster presentation, or both. Some respondents described a capstone that consisted of attending others' research presentations (e.g. departmental colloquia/seminars) and reporting on them. Among departments without a specific capstone *requirement*, an optional capstone or thesis was often required only for graduation with honors. Programs without a formal capstone option often cited limited faculty resources as the reason for the lack of a capstone, though many such programs indicated that they still saw many students engaging in research voluntarily.

Survey responses to Question 37 probing retention rates were separated from the general survey responses and scrambled in the public data to preserve anonymity. They reveal a wide range of completion rates among departments, summarized in Table 6, though programs with quite low (below 50%) rates were comparatively rare (<10%):

**Table 6.** Summary of retention rate statistics from survey question 37.

| Estimated Completion Rate | Approximate Count of Institutions | Notes |
|---|---|---|
| **90–100%** | ~20 | High retention; some programs allow switching from more rigorous to less rigorous tracks (e.g., Astrophysics → Astronomy). |
| **75–89%** | ~15 | Attrition mainly in first two years; students drop due to difficulty or interest in other majors. |
| **50–74%** | ~15 | Common issues include early math/physics hurdles and perceptions of limited career options. |
| **<50%** | ~6 | Often attributed to underprepared students or misleading expectations about the major. |
| **Unknown / Too Early to Tell** | ~10 | Newer programs or departments lacking data. |

The most commonly cited **barriers** to completion were centered on a lack of math and/or physics preparation.

Survey participants provided a broad range of responses to the open-ended final question on the survey, Question 38, which is also separated from the rest of the results in the public survey to preserve anonymity. The responses did, however, reveal two themes: best practices and challenges/needs, which are summarized in the list below with examples:

- Best practices
    - Writing-intensive courses
    - Flexible electives and curriculum pathways
    - Research experiences for students
    - Strong advising and mentoring cultures, with a focus on community-building

- Active learning and research-backed teaching practices
- Challenges
    - Enrollment growth
    - Too few faculty for rising numbers of majors
    - Shortage of research opportunities at many institutions
    - Gaps in programming experience; strong need to embed coding skills earlier in the curriculum
    - Concerns that research publications are becoming "unofficially required" for grad school admissions.

# Appendix B - Minimum Suggested Degree Requirements

The AAS Education Subcommittee on UndeRgraduate and Graduate Education (SURGE) endorses the following guidance for departments that offer, or are considering offering, an undergraduate degree in Astronomy and/or Astrophysics.

We note that the departments we surveyed to develop these recommendations span many institution types, and the *number* of courses comprising a major varies widely across them. The recommendations below are intended as a ***minimum*** set of knowledge and competencies that we believe ***all*** programs should meet. They are compiled from the courses most frequently required by the programs responding to our survey and from patterns in our data on desired and formal learning outcomes.

These recommendations are not, nor do we believe they should be, a map of the coursework required for admission to graduate programs in Astronomy and Astrophysics. Students intending to pursue graduate studies must receive clear guidance from their departments about additional coursework that they should pursue to be adequately prepared for graduate school, and graduate programs should also be clearer about these requirements (see our recommendation 10).

Our aim here is rather to provide recommendations that reflect an appropriate combination of rigour and breadth, communicating to potential employers the shared knowledge and competencies of an undergraduate degree program in Astronomy or Astrophysics, while acknowledging that many, indeed most, Astronomy majors will not pursue graduate school or careers in research.

Suggested minimum **content** requirements:

- An introductory differential and integral calculus sequence
- An introductory physics sequence consisting of, at a minimum, coverage of mechanics and electricity and magnetism, but ideally also covering some optics and waves
- At least one intermediate physics course focused on basic quantum mechanics, and ideally covering some thermal/statistical physics and relativity
- At least one astronomy/astrophysics course focused on the nature of stars, including their structure and evolution
- At least one course with a heavy emphasis on computational and data-based techniques that includes specific work with astronomical datasets, ideally within the first 2 years
- At least one course focused on observational/laboratory techniques, ideally involving some work collecting and analyzing astronomical data with telescopes

We further recommend the following minimum **competency** considerations. We note that in many cases, these could be woven into existing courses and do not necessarily need to be offered as standalone courses focused specifically on one competency. We recommend at least one course:

- in which students engage with primary astronomical literature
- with a substantial scientific writing component
- with a substantial presentation/scientific communication component
- in which students engage in collaborative, project-based work
- in which students are introduced to and/or apply advanced mathematical techniques such as multivariable calculus, differential equations, and linear algebra

We also recommend that departments consider, wherever possible, achieving depth in the astronomical applications of physical principles by requiring coursework or offering electives in:

    - Galaxies and/or cosmology
    - Exoplanets and/or planetary science
    - Radiative processes and/or the interstellar medium

# Appendix C - Sample Learning Goals

In this appendix, we provide recommendations and examples for thirteen undergraduate major-level learning goals. These match the themes outlined in Appendix A, Section 4.2 with one exception. We combine the themes "Industry, Career, or Professional Skills or Knowledge" and "Preparation for Graduate Work" into a single recommended learning goal here in order to normalize and acknowledge the many career paths that students can take post-graduation.

We also separate these learning goals into three themes - content knowledge goals, general scientific skills, and professional preparation.

**Content-Based Skills and Proficiencies**

1. ***Physics Knowledge and Skills***
   *Students will be able to apply physical principles from classical mechanics, quantum mechanics, special relativity, statistical mechanics, electromagnetism, and optics*[5] *to model and solve astrophysical problems. (Adapted from Brigham Young University)*
2. ***Astronomy and Astrophysics Knowledge and Skills***
   *Students will be able to qualitatively and quantitatively describe the contents of the extraterrestrial universe, including the physical properties, formation, and evolution of planets, stars, galaxies, and the large-scale structure of the universe itself. (Adapted from Haverford College)*
3. ***Observational, Experimental, Instrumentation or Laboratory Skills***
   *Students will be able to apply skills and knowledge related to the observational techniques, instrumentation, and computational methods used to investigate modern astrophysical phenomena and problems. (Adapted from University of Arizona)*
4. ***Data, Computational, Programming Skills***
   *Students will be able to write original computer code to accomplish computational tasks, including exploring and analyzing data, displaying astronomical images, performing calculations, and designing data visualizations. (Adapted from Penn State)*
5. ***Math, Stats, or Quantitative Knowledge and Skills***
   *Students will be able to apply quantitative skills in hypothesis formation and testing, data analysis, error estimation, conceptual and analytical modeling, and numerical computation to astrophysical problems (Adapted from Smith College)*

**General Scientific Thinking Skills**

6. ***Problem Solving or Critical Thinking Skills***
   *Students will be able to identify, formulate, and solve tractable scientific and technical problems by placing them in context, making appropriate estimates and simplifications, modeling the*

---

[5] We note that, though all Astronomy and Astrophysics majors in our survey require standalone mechanics and electromagnetism coursework, not all require modern physics, statistical mechanics/thermodynamics, or optics. As such, for majors that do not require these courses, many of their core principles can be introduced explicitly in other courses, for example quantum mechanics and statistical physics principles in a stars course or optics in an observational astronomy course. We recommend that departments that do not require additional physics coursework beyond an introductory sequence pay particular attention to where and when to introduce core advanced physics concepts into astronomy and astrophysics coursework.

*important physical processes, quantifying predictions with analytic and computational tools, and testing the correctness of the results. (Wellesley College)*

7. ***Communication Skills***
   *Students will develop proficiency with communicating, translating and interpreting fundamental astronomical concepts or research results in oral and written formats. (University of Massachusetts, Amherst)*

8. ***Collaboration Skills***
   *Students will demonstrate the ability to work collaboratively in interdisciplinary groups to solve scientific problems (Arizona State University)*

9. ***Scientific Reasoning, Method, or Process Skills***
   *Students will demonstrate scientific reasoning skills by being able to describe the roles of theory, hypothesis, and experiment in the scientific method and assess cause and effect in physical systems by formulating evidence-based logical arguments. (Adapted from University of Wyoming and University of Toledo)*

10. ***Scientific Ethics or Inclusion Skills or Knowledge***
    *Students will be able to recognize and describe the role that historic and social factors play in the practice of physics and astronomy, as well as the ways physics and astronomy have influenced history and society. (Adapted from Amherst College)*

**Professional Skills and Knowledge**

11. ***Industry, Career, and Professional Skills or Knowledge***
    *Students will generate fluency in the scientific enterprise, develop awareness of possible career paths available to the undergraduate astronomy/astrophysics majors, and hone analytical abilities and computing skills useful for careers outside of professional astronomy (Adapted from University of Hawaii at Manoa and Wesleyan University)*

12. ***Research Experience and Skills***
    *Demonstrate proficiency and independence in research techniques and methodology, including the ability to confront the unknown, generate new knowledge, and place it in context relative to what has been done before (Adapted from Rice University)*

13. ***Reading Academic Literature or Finding Scientific Information***
    *Students will demonstrate proficiency in reading and critically evaluating scientific literature and will be able to describe a question at the frontier of an astronomical discipline. (Adapted from University of Wisconsin Madison and CU Boulder)*

# Appendix D - List of Responding Institutions

## Responding Institutions with Undergraduate Degrees in Astronomy or Astrophysics

Alfred University
Amherst College
Arizona State University
Barnard College
Baylor University
Brigham Young University
California Institute of Technology
Case Western Reserve University
Colgate University
College of Charleston
Colorado College
Columbia University
Dartmouth College
Elon University
Harvard University
Haverford College
Indiana University
Lehigh University
Lycoming College
Macalester College
Montana State University
New Mexico Tech
Penn State University
Princeton University
Rice University
Rutgers University - New Brunswick
Saint Anselm College
Smith College
Sonoma State University
Stony Brook University
Swarthmore College
The Ohio State University
The University of Florida
The University of Texas at Austin
Truman State University
University of Arizona
University of Arkansas/Fayetteville
University of California, Berkeley
University of California, San Diego
University of Chicago
University of Colorado Boulder

University of Florida
University of Hawaii at Manoa
University of Illinois Champaign Urbana
University of Iowa
University of Maryland
University of Massachusetts, Amherst
University of Michigan
University of Minnesota
University of New Mexico
University of North Carolina at Chapel Hill
University of North Texas
University of Pennsylvania
University of Southern California
University of Toledo
University of Utah
University of Washington - Seattle
University of Wisconsin- Madison
University of Wyoming
Vassar College
Washington University in St. Louis
Wellesley College
Wesleyan University
Whitman College
Williams College
Youngstown State University

## Responding Institutions **without** Undergraduate Degrees in Astronomy or Astrophysics

Air Force Institute of Technology Kaduna
College of DuPage
Drexel University
Johns Hopkins University
Le Moyne College
Massachusetts Institute of Technology
New Mexico State University
Occidental College
Rowan University
Stanford University
University of South Carolina
Winona State University

AAS Education Subcommittee on UndeRgraduate and Graduate Education (SURGE)

# Report on the Landscape of Undergraduate Degrees in Astronomy and Astrophysics

**78**
Departments Surveyed

Representing
**~1000**
Undergraduate Majors Annually

YOU'RE INVITED

## AAS 247 Special Session

Monday, January 5 2:00-3:30pm
Phoenix Convention Center, 226B

come learn about and discuss the report!

Report

SCAN ME

**Motivating Questions**

- What constitutes an Astronomy/Astrophysics major? What elements are shared across programs? How much variety is there?
- What are departments preparing their students for with Astronomy or Astrophysics majors? Graduate school? Jobs in industry? Education/outreach?
- What can/should an undergraduate Astronomy/Astrophysics major communicate to graduate admissions committees and prospective employers? (i.e. what's the point?)
- What is the basic set of skills and knowledge that all Astronomy/Astrophysics majors share?
- How do undergraduate programs' goals and needs vary?

# Findings

1 Opportunities to major or concentrate in Astronomy/Astrophysics span many types and sizes of department and institution, and are offered as BA and BS degrees in nearly equal proportion (49.6% and 50.4%, respectively) .

2 Only 34% of the departments surveyed exist as separate Departments of Astronomy and/or Astrophysics. Most Astronomy/Astrophysics degrees are offered in departments with "physics" in the name (43% in Departments of Physics and Astronomy and 15% in Departments of Physics).

3 85% of departments that completed the survey offered an undergraduate degree or concentration in Astronomy and/or Astrophysics, representing a total of 66 departments and an estimated total of 718-1007 undergraduate majors annually. This is in keeping with current AIP statistics (982 bachelor's degrees awarded in the 2023-24 academic year), indicating that we have captured a large fraction of all undergraduate degree-granting programs in the United States.

4 There are a roughly equivalent number of degrees titled "Astronomy" (36%) and "Astrophysics" (30%) among the survey sample. The number of courses required for "Astrophysics" majors is, on average, *slightly* higher (21±6) than the number of courses required for "Astronomy" majors (18±5). Similarly, the number of courses required for a BS (22 ±5) is *slightly* higher than for a BA (15 ± 3). However, neither difference was significant.

5 Although "Astronomy" and "Astrophysics" majors across the country do not differ significantly in the total number of courses that they require, the terms have different popular interpretations, with "astrophysics" being perceived as more rigorous than "astronomy".

6 Differential calculus, integral calculus, introductory mechanics, and introductory electricity and magnetism are the ***only*** courses required by ***all*** Astronomy/Astrophysics degree programs.

7 Roughly ⅔ of surveyed departments indicated that their department has formal learning goals for their Astronomy/Astrophysics major. A majority (52%) indicated that their department analyzes progress toward those learning goals at least annually.

8 Survey responses show broad agreement on the core competencies that undergraduate Astronomy and Astrophysics majors should develop, but also reveal gaps between perceptions of ideal program outcomes ("informal" learning goals) and formal learning objectives. The same key skills/competencies emerge from analyses of informally-expressed learning goals, but differ markedly in the rates at which such skills are mentioned, suggesting opportunities to better integrate formal program-level learning goals with field-based expectations and norms.

9 For most programs, there is a wide range in the rates of student engagement in research and plans to attend graduate school. In other words, Astronomy and Astrophysics programs contain a highly heterogeneous mix of student experiences and goals. It is rare for a department to consist of students with homogeneous interests and paths.

# Recommendations

1 Astronomy/Astrophysics majors show healthy enrollments across a wide range of institution, department, and degree types. Departments that do not currently offer an astronomy-specific credential should adopt a major or concentration in Astronomy/Astrophysics.

2 Our data do not support a distinction between "Astronomy" and "Astrophysics" degrees. Departments should adopt the term "astrophysics" to describe their academic majors *wherever it is feasible to do so.*

3 The AAS Board of Trustees should endorse a set of recommended course requirements for Astronomy/Astrophysics undergraduate degree-granting programs. ***Our recommended requirements are listed in Appendix B of the report.***

4 All academic departments should adopt formal program-level learning goals that are well-aligned with expected outcomes for students graduating from the program. These should encompass the content knowledge and skills that they believe ***all*** majors should leave the program with. These learning goals should map to specific (ideally multiple) courses within the major requirements and should be phrased in such a way that they can be easily assessed, both by the department/institution internally and externally by departmental reviewers or accreditors. ***Examples of well-phrased learning goals are listed in Appendix C of the report.***

5 Given the high rate at which astronomers who responded to our survey specified data analysis and scientific computing skills among the most important things majors should leave Astronomy/Astrophysics programs with, we especially strongly recommend that all programs that do not currently have formal learning goals in this area adopt them.

6 Departments should work to embed computational competencies as early as possible into the coursework for Astronomy and Astrophysics majors. We recommend that the AAS and its affiliate committees place a high priority on the development of pedagogical resources aimed at teaching scientific computing in Astronomy.

7 That the AAS sponsor a virtual or in-person meeting of the existing "Astronomy Department Chairs and Program Directors" group with the goal of revising as necessary and endorsing the recommendations of the report.

8 That the AAS create and maintain its own email list of department chairs. This list should be used sparingly to support transmission of key information from AAS committees and leadership that is especially relevant at the department level.

9 Care should be taken to ensure that any course recommendations and learning goals endorsed by the AAS for Astronomy/Astrophysics majors are relevant across the broad spectrum of careers that Astronomy/Astrophysics majors pursue, and are not narrowly focused on graduate preparation.

10 The AAS Working Group on Graduate Admissions should work to develop a more complete list of required/recommended knowledge and skills/competencies for graduate school and advertise it broadly to both undergraduate programs and early career researchers. *We do not endorse a move toward all undergraduate astronomy curricula fulfilling all grad school requirements* (see recommendation 9), but believe that undergraduate programs should consider these needs in planning and advertising their curricula and elective course offerings, as well as advising undergraduates who intend to pursue graduate studies.

Feedback

SCAN ME

Read the full report
https://bit.ly/AAS_SURGE_report

And provide feedback

Report

SCAN ME